\documentclass[prr, 10pt, twocolumn, tightenlines, aps, longbibliography]{revtex4-1}

\usepackage[utf8]{inputenc}
\usepackage[T1]{fontenc}
\usepackage[paper=a4paper,left=20mm,right=20mm,top=25mm,bottom=20mm]{geometry}
\usepackage{float}
\usepackage[english]{babel}

\usepackage{mathtools}
\usepackage{graphicx}
\usepackage{multirow}
\usepackage{tabularx}
\usepackage[sgvnames, table]{xcolor}
\usepackage{soul}
\usepackage[colorlinks=true,allcolors=blue]{hyperref}

\DeclarePairedDelimiter\bra{\langle}{\rvert}
\DeclarePairedDelimiter\ket{\lvert}{\rangle}
\DeclarePairedDelimiter\expval{\langle}{\rangle}
\DeclarePairedDelimiterX\braket[2]{\langle}{\rangle}{#1 \delimsize\vert #2}

\begin{abstract}
	We unravel the many-body dynamics of a harmonically trapped impurity colliding with a bosonic medium confined in a double-well upon quenching the initially displaced harmonic trap to the center of the double-well. We reveal that the emerging correlation dynamics crucially depends on the impurity-medium interaction strength allowing for a classification into different dynamical response regimes. For strong attractive impurity-medium couplings the impurity is bound to the bosonic bath, while for intermediate attractions it undergoes an effective tunneling. In the case of weak attractive or repulsive couplings the impurity penetrates the bosonic bath and performs a dissipative oscillatory motion. Further increasing the impurity-bath repulsion results in the pinning of the impurity between the density peaks of the bosonic medium, a phenomenon that is associated with a strong impurity-medium entanglement. For strong repulsions the impurity is totally reflected by the bosonic medium. To unravel the underlying microscopic excitation processes accompanying the dynamics we employ an effective potential picture. We extend our results to the case of two bosonic impurities and demonstrate the existence of a qualitatively similar impurity dynamics.
\end{abstract}

\begin{document}
\title{Many-body collisional dynamics of impurities injected into a double-well trapped Bose-Einstein condensate}
\author{Friethjof Theel $^1$}
\author{Kevin Keiler $^1$}
\author{Simeon I. Mistakidis $^1$}
\author{Peter Schmelcher $^{1,2}$}
\affiliation{$^1$Center for Optical Quantum Technologies, University of
	Hamburg, Department of Physics, Luruper Chaussee 149, D-22761, Hamburg, Germany}
\affiliation{$^2$The Hamburg Centre for Ultrafast Imaging, University of
	Hamburg, Luruper Chaussee 149, D-22761, Hamburg, Germany}

\maketitle

\section{Introduction}

Due to their extraordinary controllability ultracold atoms have been used to study various properties of many-body quantum systems. Indeed, they can be confined in arbitrary trapping geometries and dimensions \cite{henderson2009, ketterle1996, gorlitz2001, petrov2004}, the underlying interatomic interactions are tunable via Feshbach resonances \cite{fano1961, kohler2006, chin2010, fedichev1996, inouye1998} while mixtures of quantum gases namely Bose-Bose \cite{pflanzer2009, jorgensen2016}, Bose-Fermi \cite{stan2004, shin2008} and Fermi-Fermi ones \cite{wille2008, kohstall2012, koschorreck2012, scazza2017} can be realized. Recently, major attention has been placed on strongly particle imbalanced mixtures where for instance a single impurity is immersed in a many-body environment. Here, the concept of a polaron \cite{landau1933}, which has been exhaustively studied in solid state physics, can be recovered where the impurity plays the role of an effective particle dressed by the excitations of its surrounding. In this context, the existence and characteristics of Fermi \cite{nascimbene2009, ngampruetikorn2012, massignan2014, schmidt2018, gamayun2018} and Bose polarons \cite{palzer2009, fukuhara2013, volosniev2017, dehkharghani2018, mistakidis2019c, albiez2005, ardila2019, skou2021} have been unveiled, mainly focusing on their stationary properties \cite{privitera2010, kain2016, grusdt2017a, ichmoukhamedov2019, albiez2005, ardila2019a, ardila2020} and more recently on the dynamics \cite{boyanovsky2019, kamar2019, mistakidis2020, bouton2020} of these quasi-particles.

The involved confining potentials have a major impact on the dynamical behavior of the impurity. For instance it has been shown that the impurity-medium interaction quench dynamics in a harmonic trap leads to oscillatory \cite{volosniev2015}, dipole-like and dissipative impurity \cite{mistakidis2019} motion depending on the impurity-bath coupling strength or to temporal orthogonality catastrophe events for strong repulsions \cite{mistakidis2019c}. Another important aspect of such impurity settings concerns their transport properties through the environment \cite{nielsen2019}. Indeed, the tunneling dynamics of impurities confined in a double-well and coupled to a lattice trapped medium has been studied in the context of an effective potential \cite{theel2020, keiler2020}. Additionally, dephasing and clustering processes \cite{klein2007, keiler2018a} as well as distinct transport pathways \cite{keiler2019} were observed for impurities confined in lattice potentials. Furthermore, the collisional dynamics of impurities with a Bose-Einstein condensate (BEC) has been studied experimentally \cite{schmidt2018a, schmidt2019a} and theoretically \cite{lingua2018}. In the latter case the complete reflection of the impurities from a harmonically trapped BEC, their trapping within the bath as well as the generation of dark and bright solitons have been revealed in the absence of correlations.

Apart from the above-described intriguing collisional channels, certainly a much richer dynamical response is expected to emerge in the presence of a lattice potential. Here, the periodic structure of the medium's density imprinted by the external potential acts as a material multi-barrier which enforces specific tunneling pathways for the impurity. A minimal setup of this type consists of a bosonic bath trapped in a double-well potential, where complex dynamical response regimes of the impurity are anticipated. For instance, dephasing dynamics of the impurity associated with enhanced energy redistribution processes can be triggered \cite{kronke2015a, mistakidis2019} or in the case of attractive impurity-bath coupling strengths bound states can emerge. Moreover, the back-action of the impurity on the bosonic background where the former is expected to induce tunneling of the medium emulating a Josephson junction \cite{julia-diaz2009} is certainly of interest. Here, also the intraspecies coupling of the bath particles enforcing the latter to configure in a superfluid or Mott-state, thus affecting their mobility, is expected to impact the impurity's response. In this context the interplay of the boson-boson interaction with the impurity-impurity induced correlations is also a relevant direction of study. Another interesting aspect is whether the emergent dynamical response regimes, found in the double-well case, remain robust in a setup where the bath is trapped in a multi-well potential. Additionally, due to the collision of the impurities with their environment strong impurity-medium correlations are expected to emerge giving rise to beyond-mean-field collisional channels. To trace the non-equilibrium quantum dynamics, we employ the Multi-Layer Multi-Configuration Time-Dependent Hartree method for atomic mixtures (ML-MCTDHX) \cite{cao2013, kronke2013, cao2017} which is capable of capturing all relevant inter- and intraspecies correlations.

To address these aspects we consider a harmonically trapped impurity which is coupled via a contact interaction potential to a bosonic environment confined in a double-well. The dynamics is induced by quenching the initially displaced harmonic confinement of the impurity to the center of the double-well. By steering the impurity-medium interaction strength from attractive to strongly repulsive values we are able to identify five dynamical response regimes in the case of a single impurity \cite{mukherjee2020}. These regimes range from a bound state formation between the impurity and its environment for strong attractive impurity-bath interaction strengths to its dissipative oscillatory motion \cite{knap2014, mistakidis2019} within the bosonic background at weak attractive and repulsive couplings and finally its total reflection from the medium for strong repulsive interactions. For intermediate attractive or repulsive interaction strengths the impurity effectively tunnels between the sites of the double-well potential or it is pinned between the later, respectively \cite{theel2020}. In all of the above-mentioned cases we reveal the build up of a significant impurity-medium entanglement \cite{mistakidis2019c} which is mostly pronounced in the dissipative oscillation and the pinning regimes. To unravel the microscopic processes participating in the dynamics we construct an effective potential \cite{mistakidis2019, mistakidis2019a, keiler2019, mistakidis2020}. This picture enables us to understand the dynamical behavior of the impurity in all response regimes and, in particular, uncover hidden excitations in the pinning regime.
Extending our results to the two-impurity case we identify five qualitatively similar response regimes as compared to the single-impurity scenario. In this case we explicate the involvement of single- and two-particle excitation processes of the impurities within the effective potential \cite{erdmann2019} and also reveal the interplay of impurity-impurity induced correlations for different intraspecies interactions of the bath.
To demonstrate the generalization of the identified dynamical response regimes of the impurity we additionally consider a bosonic bath trapped in a triple-well. In this context we find that the steady bound state, the dissipative oscillation and the total reflection regimes remain robust, see in particular Appendix \ref{ap:triple_well}.

This work is structured as follows. In section \ref{sec:setup} we introduce the system under investigation and specify the used quench protocol. The employed variational method to trace the many-body dynamics is outlined in section \ref{sec:ML-X}. Section \ref{sec:dyn_reg_single_imp} provides a detailed classification and analysis of the dynamical response regimes in dependence of the impurity-medium interaction strength. We extend our results to two impurities in section \ref{sec:dyn_reg_two_imp} and conclude this work in section \ref{sec:summary_outlook} providing a summary and an outlook of possible future research directions. In the appendices we further elaborate on the features of the identified dynamical response regimes discussing energy redistribution processes (appendix \ref{ap:energy}), the impurity-medium two-body correlation dynamics at strong attractions (appendix \ref{ap:steady_bound_state_regime}), the effective mass of the impurity (appendix \ref{ap:diss_osc_reg}) and the exposure of hidden excitations revealed for repulsive impurity-medium couplings (appendix \ref{ap:pinning_regime}). Appendix \ref{ap:triple_well} demonstrates the collisional dynamics when considering a bosonic bath in a triple-well.

\section{Setup and quench protocol}
\label{sec:setup}

Our setup consists of two different species of bosons $B$ and $I$, also referred to as the medium and impurity species, respectively. For the two species we consider $N_B$ and $N_I$ particles of mass $m_B$ and $m_I$, respectively. We operate in the ultracold regime and thus s-wave scattering is the dominant process allowing us to model the interaction between the atoms with a contact interaction potential \cite{olshanii1998}. Therefore, we employ for the impurity-bath interaction a contact interaction potential of strength $g_{BI}$, while particles of the same species interact with a contact interaction potential among each other with strengths $g_{BB}$ for the environment and $g_{II}$ for the impurity species. Each species is confined in a different one-dimensional optical potential $V_\sigma$ at zero temperature. This can be easily achieved experimentally \cite{leblanc2007, lercher2011, jorgensen2016, hu2016} especially for the mass-imbalanced case under consideration of, i.e. $\prescript{87}{}{\text{Rb}}$ atoms for the bath and $\prescript{133}{}{\text{Cs}}$ atoms for the impurity species. The resulting Hamiltonian of the system reads $\hat{\mathcal{H}} = \mathcal{\hat{H}}^B + \mathcal{\hat{H}}^I + \mathcal{\hat{H}}^{\textrm{int}}$ where
\begin{align}
\mathcal{\hat{H}}^\sigma =& \sum_{i=1}^{N_\sigma}
\bigg( - \frac{\hbar^2}{2m_\sigma} \frac{\partial^2}{(\partial x_i^{\sigma})^2}	+ V_\sigma(x_i^\sigma) \nonumber \\
&+ g_{\sigma\sigma} \sum_{i<j} \delta(x_i^\sigma-x_j^\sigma) \bigg)
\end{align}
is the Hamiltonian of species $\sigma\in\{B,I\}$. The bosonic medium and impurity species are coupled via $\mathcal{\hat{H}}^{\textrm{int}} = g_{BI} \sum_{i=1}^{N_B}\sum_{j=1}^{N_I}\delta(x_i^B-x_j^I)$.
The impurities are confined in a harmonic oscillator potential $V_I(x_i^I) = m_I\omega_I^2(x_i^I + x_0^I)^2/2$, where $\omega_I$ is the trapping frequency with $x_0^I$ being the spatial displacement of the trap. The environment is trapped in a double-well $V_B(x_i^B) = m_B\omega_B^2(x_i^B)^2/2 + \frac{h_B}{\sqrt{2\pi}w_B} \exp\!\left(-\frac{(x_i^B)^2}{2(w_B)^2}\right)$ which is constructed by superimposing a harmonic oscillator potential of frequency $\omega_B$ and a Gaussian of width $w_B$ and height $h_B$ \cite{albiez2005}.
We consider for the bosonic medium $N_B=20$ $\prescript{87}{}{\text{Rb}}$ atoms with mass $m_B=1$, and for the impurity species $N_I=1,2$ $\prescript{133}{}{\text{Cs}}$ atoms with mass $m_I=133/87$ \cite{haas2007}.

Experimentally, a one-dimensional potential can be realized by employing a strong harmonic confinement along the transverse direction in order to freeze out the relevant degrees of freedom \cite{olshanii1998, katsimiga2020}. Subsequently, we provide the energy of the Hamiltonian $\hat{\mathcal{H}}$ in terms of $\tilde{E}=\hbar\tilde{\omega}$, where $\tilde{\omega}$ is the frequency of the perpendicular confinement. The length and time scales are then expressed in units of $\tilde{x}=\sqrt{\hbar/(m_B\tilde{\omega})}$ and $\tilde{\omega}^{-1}=\hbar \tilde{E}^{-1}$, respectively. Regarding the frequency of the harmonic oscillator potential of the impurities we use $\omega_I/\tilde{\omega}=0.2$. For the harmonic contribution to the double-well potential we employ a frequency of $\omega_B/\tilde{\omega}=0.15$ and for the Gaussian a width of $w_B/\tilde{x}=0.8$ and a height of $h_B/\tilde{E}\tilde{x}=2.0$ leading to a central barrier of the double-well below which six energetically lowest eigenstates of the corresponding one-body Hamiltonian are located.

\begin{figure}[t]
	\begin{center}
		\includegraphics[width=.45\textwidth]{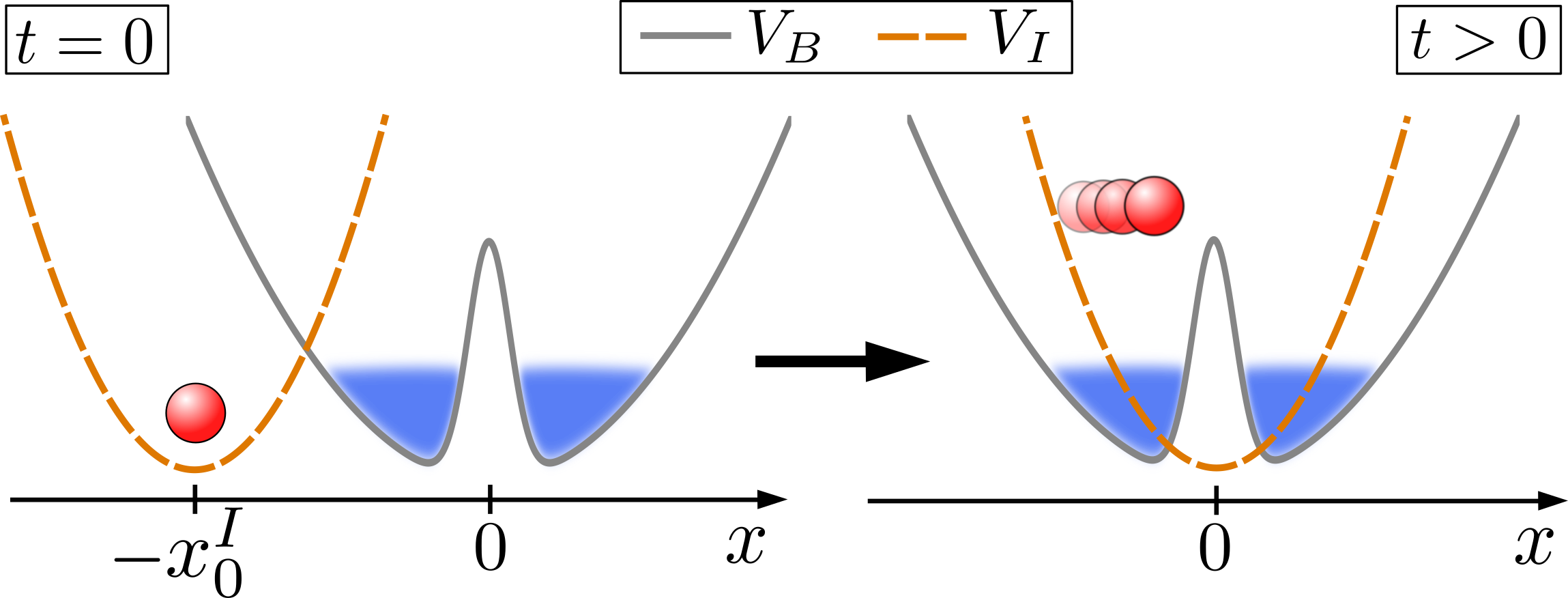}
		\caption{Sketch of the setup under consideration at $t=0$ (left panel). The harmonic trap of the impurity (red circle) is initially displaced by $x_0^I$ with respect to the center of the double-well potential of the bosonic medium (blue shaded area). The dynamics is induced by quenching the potential of the impurity to $x_0^I=0$ (right panel).}
		\label{fig:hoc_setup}
	\end{center}
\end{figure}

We prepare the system in its ground state with the harmonic trap of the impurities displaced by $x_0^I$. The spatial overlap between the species in the non-interacting case is of about 3.5\,\% for $x_0^I=8$ and increases (decreases) for attractive (repulsive) impurity-medium interactions. After a $g_{BI}$-dependent ground state is found the dynamics is induced by quenching, at $t=0$, the trap center of the impurity's harmonic potential to the center of the bosonic environment, i.e. setting $x_0^I=0$ (see Figure \ref{fig:hoc_setup} for $t>0$). Thereby, a collision of the initially displaced impurities with the bosonic medium is triggered, a process that strongly depends on the impurity-medium coupling strength $g_{BI}$ as we shall demonstrate below.

\section{Many-body wave function ansatz}
\label{sec:ML-X}

To calculate the quantum dynamical behavior of the binary system we employ the \textit{ab initio} Multi-layer Multi-Configuration Time-Dependent Hartree method for atomic mixtures (ML-MCTDHX) \cite{cao2013, kronke2013, cao2017}. Within this approach we express the many-body wave function $\ket{\Psi^{\textrm{MB}}(t)}$ of the binary mixture using the Schmidt decomposition \cite{schmidt1907,horodecki2009}
\begin{equation}
\ket{\Psi^{\textrm{MB}}(t)} = \sum_{i=1}^{M} \sqrt{\lambda_i(t)} \ket{\Psi_i^{B}} \otimes \ket{\Psi_i^{I}}.
\label{eq:schmidt_decomp}
\end{equation}
For our purposes we expand each species in $M=6$ species functions $\ket{\Psi_i^{\sigma}}$ with $\sigma\in\{B,I\}$. Moreover, the species functions are weighted with the time-dependent Schmidt-coefficients $\lambda_i$ which contain information about the entanglement between the two species. For instance, in the case of only one non-vanishing Schmidt-coefficient the species are considered to be not entangled since the system can be described by a single product state (species mean-field ansatz) \cite{horodecki2009, roncaglia2014}. Next, each species function is expanded in a set of time-dependent number states $\ket{\vec{n}_i^\sigma (t)}$
\begin{equation}
\ket{\Psi_i^{\sigma}} = \sum_{\vec{n}|N_\sigma} C_{i,\vec{n}}^{\sigma}(t) \ket{\vec{n}_i^\sigma (t)},
\label{eq:spec_fct}
\end{equation}
with time-dependent coefficients $C_{i,\vec{n}}^{\sigma}$. Each number state $\ket{\vec{n}_i^\sigma (t)}$ determines the configurational occupation of $N_\sigma$ particles on $d_{\sigma}$ single-particle functions (SPFs) where, at the same time, the number of occupied SPFs must add up to the total particle number $N_\sigma$ (indicated by $\vec{n}|N_\sigma$). In this work we employed $d_{B}=d_{I}=6$ SPFs. Eventually, the single-particle functions are represented in a time-independent discrete variable representation (DVR) \cite{light1985}.
The propagation in time is performed by employing the Dirac-Frenkel variational principle \cite{dirac1930, raab2000} leading to a set of equations of motion for the system (see for more details \cite{cao2017, kohler2019}).

The advantage of this method is its underlying multi-layering architecture of the total wave function combined with its time-dependent basis set [Eqs. (\ref{eq:schmidt_decomp}) and (\ref{eq:spec_fct})]. Especially, with the latter a co-moving basis set is utilized leading to a significant reduction of required basis functions compared e.g. to an exact diagonalization approach. On the other hand, the multi-layering structure provides access to all relevant inter- and intraspecies correlations of the system in an efficient manner.

\section{Dynamical response regimes of a single impurity}
\label{sec:dyn_reg_single_imp}

In the following we analyze the collisional dynamics of a single impurity ($N_I=1$) trapped in a harmonic oscillator and interacting with a bosonic medium confined in a double-well. Initially, the system is prepared in its ground state with the impurity's harmonic trap being spatially shifted by $x_0^I/\tilde{x}=8$ with respect to the center of the double-well. The dynamics is induced by quenching the harmonic oscillator of the impurity to $x_0^I/\tilde{x}=0$. Varying the impurity-medium interaction strength $g_{BI}$ from the strongly attractive to the strongly repulsive regime we discuss the emergent dynamical response of the impurity and its backaction to the environment. The intraspecies interaction strength between the medium particles is fixed to $g_{BB}/\tilde{E}\tilde{x}=0.5$.

\begin{figure}[t]
	\begin{center}
		\includegraphics[width=\linewidth]{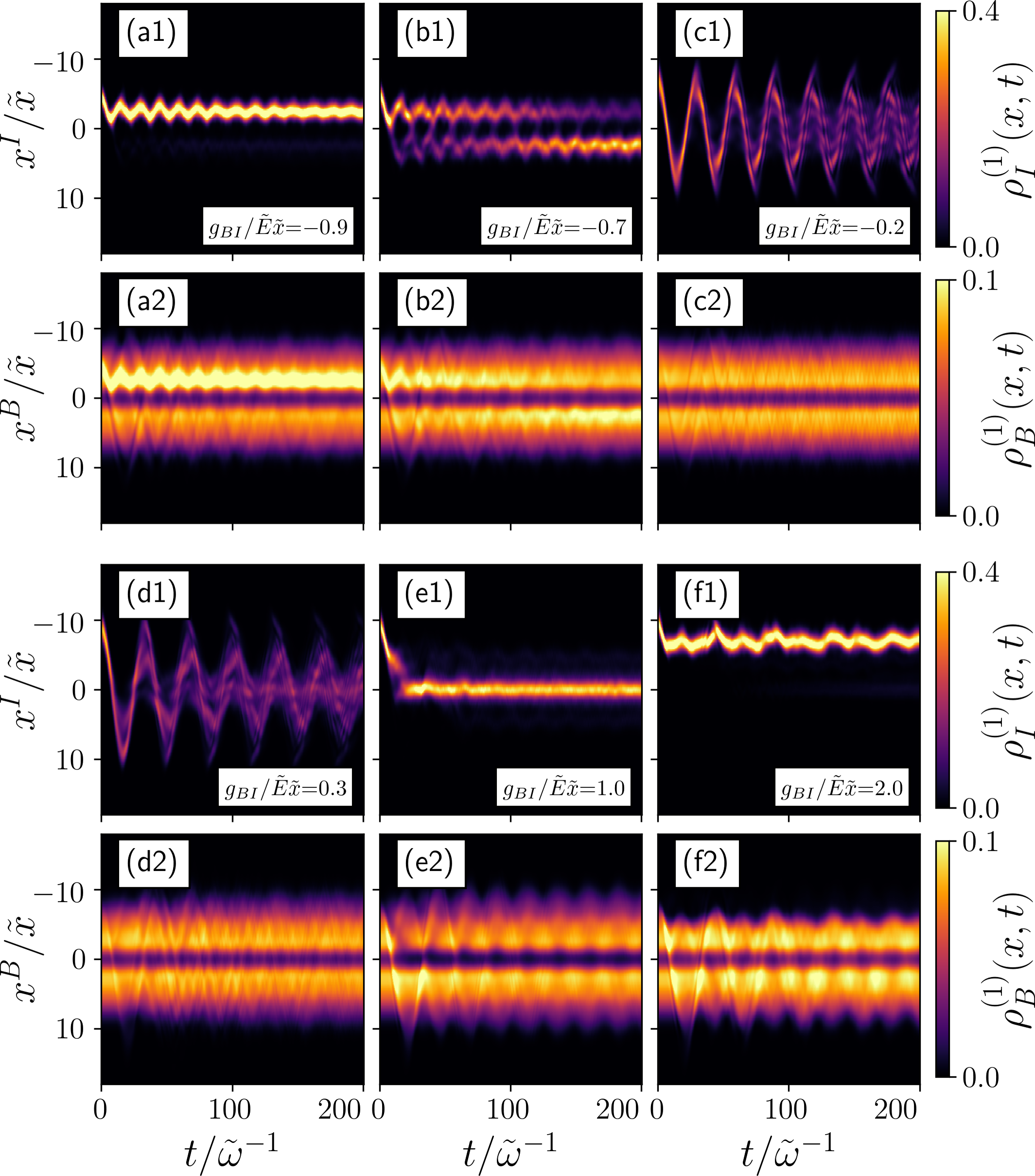}
		\caption{Temporal evolution of the one-body density of (a1)-(f1) the impurity and (a2)-(f2) the bosonic medium. We induce the dynamics by quenching the initial displacement $x_0^I/\tilde{x}=8$ of the harmonic trap to $x_0^I=0$. Different dynamical response regimes are realized by steering the impurity-medium interaction strength $g_{BI}$ (see legends). The system consists of a $N_I=1$ impurity and $N_B=20$ bath particles with $g_{BB}/\tilde{E}\tilde{x}=0.5$.}
		\label{fig:gpop}
	\end{center}
\end{figure}

To reveal the overall dynamical response of the system we initially inspect the one-body density $\rho^{(1)}_\sigma(x,t)=\bra{\Psi^{\textrm{MB}}(t)}\hat{\Psi}_\sigma^\dagger(x)\hat{\Psi}_\sigma(x)\ket{\Psi^{\textrm{MB}}(t)}$ of species $\sigma$, where $\hat{\Psi}_\sigma(x)$ is the bosonic field operator of the corresponding species. The spectral decomposition of the one-body density \cite{sakmann2008, peschel2009} reads
\begin{equation}
\rho^{(1)}_\sigma(x,t) = \sum_j n_{j}^\sigma(t) \Phi_{\sigma,j}^*(x,t) \Phi_{\sigma,j}(x,t),
\label{eq:obd}
\end{equation}
where $n_{j}^\sigma$ denote the natural populations and $\Phi_{\sigma,j}$ the natural orbitals of species $\sigma$.

\begin{figure*}
	\centering
	\includegraphics[width=0.85\linewidth]{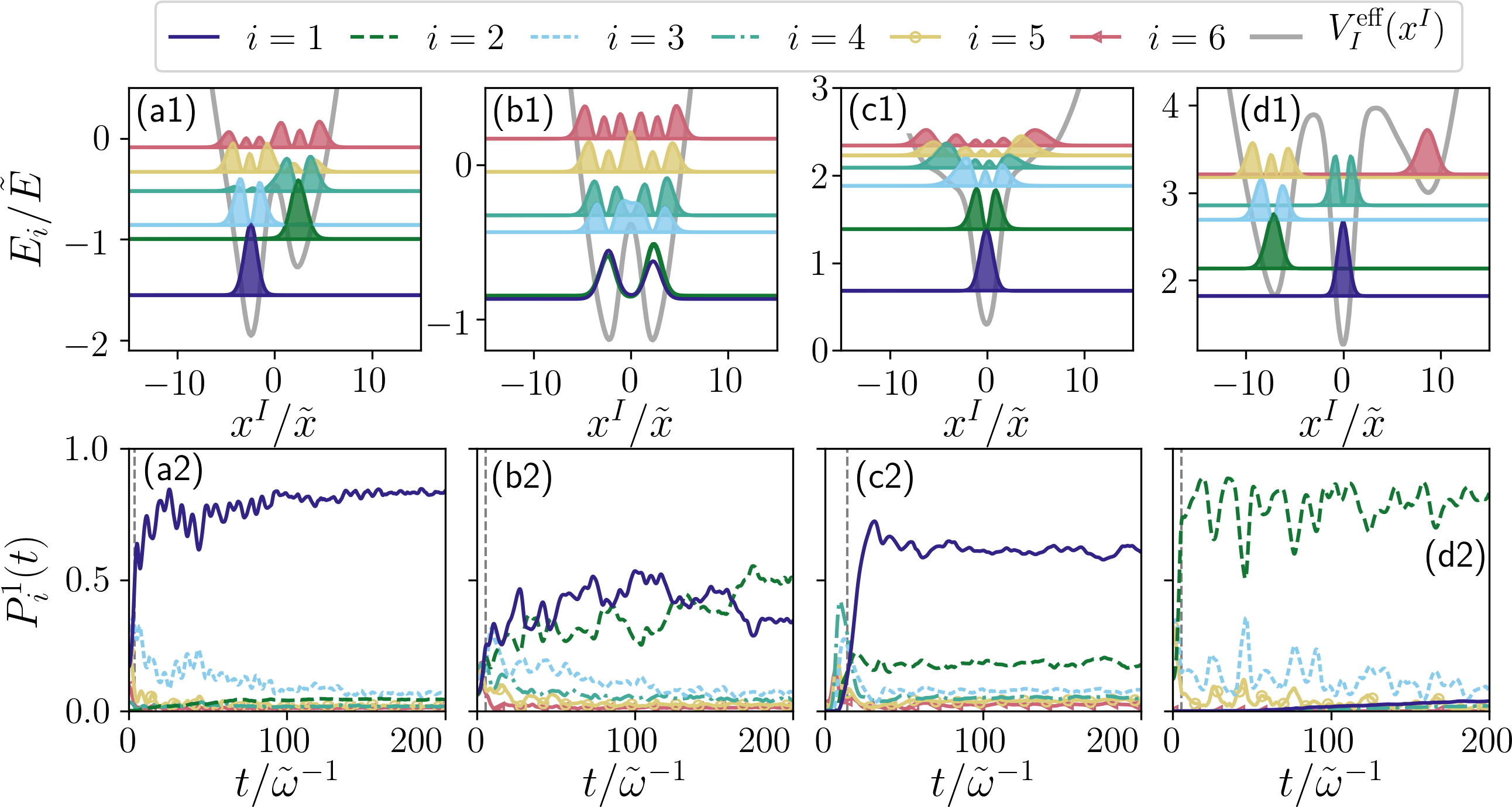}
	\caption{(a1)-(d1) Time-averaged effective potential of the impurity $V_I^{\textrm{eff}}(x^I)$ (gray solid lines) together with the first six energetically lowest eigenfunctions (see legend) shifted by the respective eigenenergies obtained from the associated single-particle Hamiltonian $\hat{\mathcal{H}}^{(1)}$ (see text). (a2)-(d2) Time-dependent probabilities for the impurity to occupy one of the eigenfunctions shown in (a1)-(d1), respectively. Each column corresponds to a particular dynamical response regime (with a specific value of $g_{BI}$) which is sorted from left to right by $g_{BI}/\tilde{E}\tilde{x}=-0.9,-0.7,1.0,2.0$. $\sum_i P_i^1(t)>0.94$ is fulfilled for times later than the threshold time depicted by the gray dashed lines.}
	\label{fig:TAEP_sOccup}
\end{figure*}

Figure \ref{fig:gpop} presents the time-evolution of $\rho^{(1)}_\sigma(x,t)$ for different impurity-medium interaction strengths $g_{BI}$ ranging from strongly attractive to strongly repulsive values.
In Figures \ref{fig:gpop}(a1) and (a2) we monitor the one-body density of the impurity $\rho^{(1)}_I(x,t)$ and the bosonic environment $\rho^{(1)}_B(x,t)$ in the course of the evolution for $g_{BI}/\tilde{E}\tilde{x}=-0.9$ namely for strong attractive couplings. As a result of the quench the impurity starts to oscillate with a small amplitude which decays during the time-evolution. Thereby, the spatial maximum of $\rho^{(1)}_I(x,t)$ remains in the vicinity of the left site of the double-well potential \footnote{We have checked that also for a longer evolution time $t/\tilde{\omega}^{-1} = 1600$ the impurity remains localized at the left site of the double-well.}. Also $\rho^{(1)}_B(x,t)$ exhibits a maximum at the same location [cf. Figure \ref{fig:gpop}(a2)], a phenomenon that is attributed to the strong impurity-bath attraction \cite{sacha2006}. Due to this behavior, i.e. the enhanced spatial localization tendency of the impurity and the medium, and the fact that the impurity dominantly occupies a state with negative eigenenergy [see for details Figure \ref{fig:TAEP_sOccup}(a1)] we refer to this dynamical response regime as the \textit{steady bound state regime}. A similar dynamical response where the impurities localize at the maximum of the medium's one-body density has also been observed in the case of a harmonically trapped bath \cite{lingua2018}.

For intermediate attractive impurity-medium interactions, $g_{BI}/\tilde{E}\tilde{x}=-0.7$, corresponding to the one-body densities shown in Figures \ref{fig:gpop}(b1) and (b2) the impurity does not localize exclusively on one site of the double-well anymore as in the above discussed case. Rather, we observe a decay of $\rho^{(1)}_I(x,t)$ on the left site and a simultaneous increase at the right site of the double-well. This response of the impurity is reminiscent of the tunneling dynamics of a single particle confined in a double-well \cite{zollner2008}. Later on, we will show that the effective potential encountered by the impurity resembles a double-well since it accounts for the effects of the attractive impurity-medium coupling [cf. Figure \ref{fig:TAEP_sOccup}(b1)]. In this sense we label this response region as the (effective) \textit{tunneling regime}. Note that the back-action of the impurity on the bosonic medium leads to a shift of the maximum of $\rho^{(1)}_B(x,t)$ following the impurity's tunneling behavior. Interestingly, this effective tunneling behavior of the impurity at intermediate attractions resembles the dynamical response of an impurity trapped in a double-well and repulsively coupled to a lattice trapped bosonic bath \cite{theel2020}.

The next response regime, which we will refer to as the \textit{dissipative oscillation regime}, emerges at weakly attractive or repulsive impurity-medium interaction strengths, e.g. for $g_{BI}/\tilde{E}\tilde{x}=-0.2,0.3$. Here, the impurity-medium coupling is sufficiently small such that the impurity is able to completely penetrate its environment. Consequently, the impurity initiates a tunneling of $\rho^{(1)}_B(x,t)$ from one site of the double-well to the other which decays in the course of time. The resulting impurity dynamics turns out to be a decaying oscillatory motion with an initial amplitude as large as the spatial extent of $\rho^{(1)}_B(x,t)$, cf. Figures \ref{fig:gpop}(c1), (c2) and (d1), (d2). We attribute this decay process to a continuous energy transfer from the impurity to the medium [see Appendix \ref{ap:energy} for details]. A more detailed analysis of this dissipative behavior estimating also the effective mass of the emergent quasi-particle can be found in Appendix \ref{ap:diss_osc_reg}. Note, that such a dissipative behavior of impurities is a generic feature caused by the build up of impurity-medium correlations and has been reported, e.g., in \cite{knap2014, lingua2018, mistakidis2019}.

In the case of intermediate repulsive impurity-bath couplings, e.g. $g_{BI}/\tilde{E}\tilde{x}=1.0$, we observe a spatial localization tendency of the impurity at the trap center accompanied by vanishing oscillations [cf. Figure \ref{fig:gpop}(e1)]. Therefore, we refer to this response regime as the \textit{pinning regime}. During the impurity's localization process, an intra-well dynamics is induced on the bosonic background. Here, the central barrier of the double-well is effectively enlarged due to the material barrier created by the impurity and leading to an oscillatory motion of the bath cloud in each site of the double-well.
Further increasing $g_{BI}$ to strong impurity-medium interaction strengths, e.g. $g_{BI}/\tilde{E}\tilde{x}=2.0$, the underlying repulsion between the impurity and the bath particles in the left site of the double-well becomes sufficiently large such that the impurity is totally reflected [see Figure \ref{fig:gpop}(f1)]. In this sense, we shall address this behavior as the \textit{total reflection regime}. Thereby, the bosonic environment experiences a population imbalance in the double-well potential [cf. Figure \ref{fig:gpop}(f2)] which is accompanied by the phase separation between the medium and the impurity, a well-known process occurring in the strongly repulsive case \cite{lingua2018, mistakidis2019c}.

In conclusion, we have captured five different dynamical response regimes of the impurity depending on the impurity-medium interaction strength. Remarkably, by tuning the impurity-bath coupling $g_{BI}$ it is possible to control the location of the impurity.
Note that, in each of the above-described regimes Josephson-like oscillations of the bath take place induced by the coupling to the impurity.
Furthermore, in all response regimes, apart from the total reflection regime, the impurity exhibits a finite spatial overlap with the bath in the course of the evolution. Thus, it can be dressed by the excitations of the latter allowing in principle for quasi-particle, in particular Bose polaron, formation \cite{rath2013, mistakidis2019c, mistakidis2020a}. In the total reflection regime the impurity phase separates from its environment after the first collision and, therefore, the polaron, even it is formed for very short evolution times, decays. Similar manifestations of a decaying polaron formation at strong repulsive impurity-medium couplings in the course of the evolution have been already reported in the literature being referred to as temporal orthogonality catastrophe events \cite{goold2011, mistakidis2019c, mistakidis2020a}.

\begin{figure}[h!t]
	\centering
	\includegraphics[width=\linewidth]{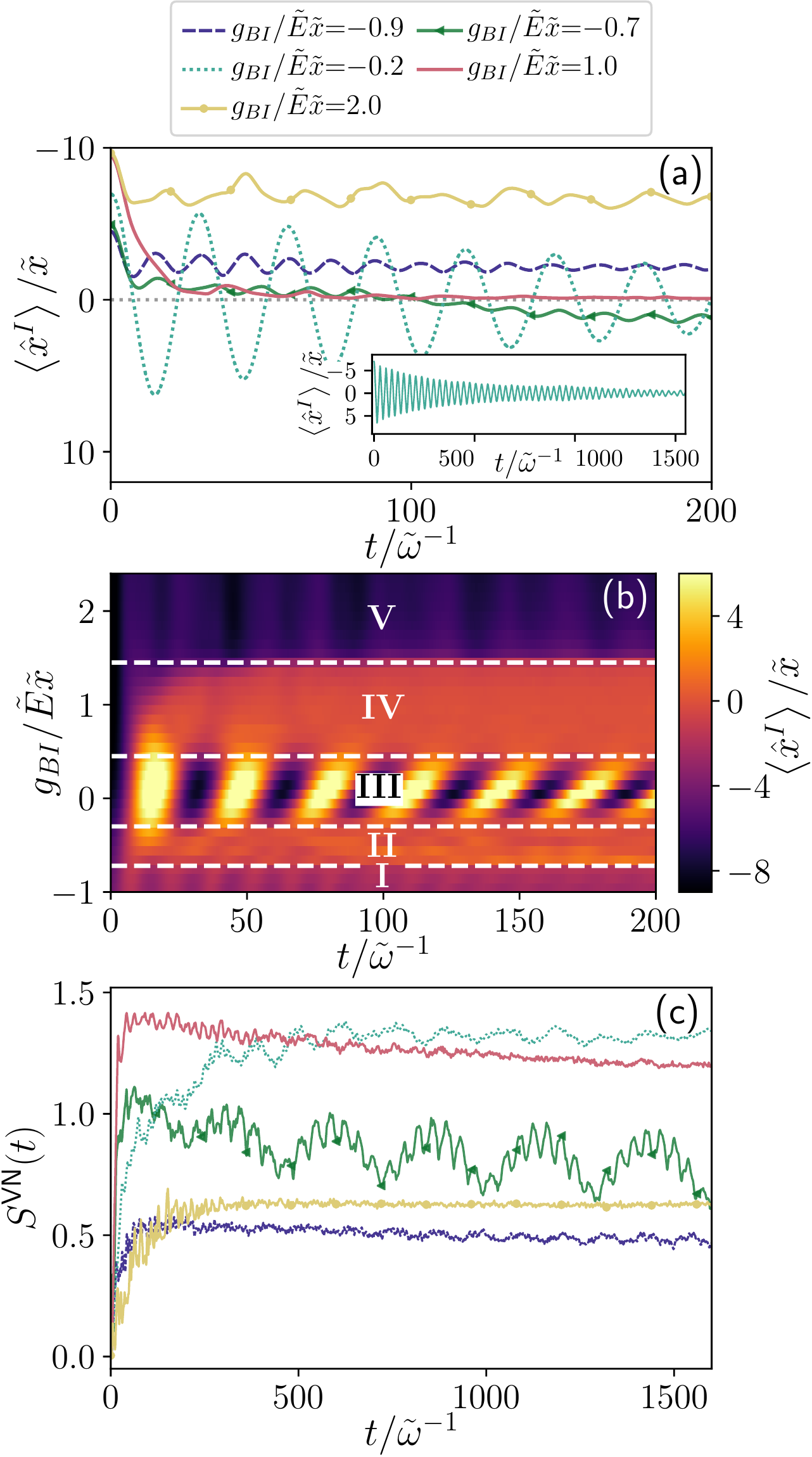}
	\caption{(a) Time-evolution of the mean position of the impurity for different impurity-medium interaction strengths $g_{BI}$ (see legend). Each value of $g_{BI}$ corresponds to one of the five dynamical response regimes of the impurity. The inset of (a) shows the long-time evolution of the mean position obtained for $g_{BI}/\tilde{E}\tilde{x} = -0.2$. (b) Temporal evolution of the mean position as a function of the impurity-medium interaction strength $g_{BI}$. (c) Long-time evolution of the von-Neumann entropy $S^{\text{VN}}(t)$ obtained for the same $g_{BI}$ as used in (a).}
	\label{fig:vNE_mp_cmp}
\end{figure}

\begin{figure*}
	\centering
	\includegraphics[width=0.9\linewidth]{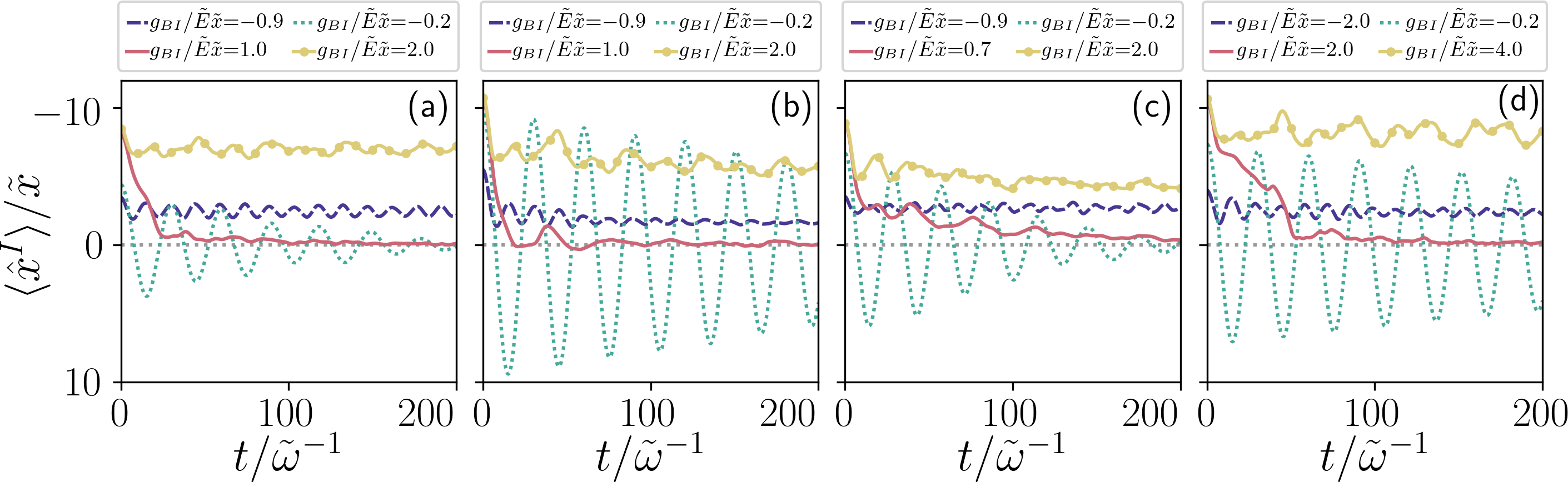}
	\caption{Time-evolution of the impurity's mean position for varying system parameters (see legends). (a) corresponds to an initial displacement $x_0^I/\tilde{x}=5$ of the impurity's harmonic trap and (b) refers to $x_0^I/\tilde{x}=10$. In both cases $g_{BB}/\tilde{E}\tilde{x}=0.5$. In panels (c) and (d) the intraspecies interaction strengths are set to $g_{BB}/\tilde{E}\tilde{x}=0.2$ and $g_{BB}/\tilde{E}\tilde{x}=1.0$, respectively, for fixed $x_0^I/\tilde{x}=8$. The impurity-medium couplings $g_{BI}$ are chosen such that the mean positions exhibit a behavior which can be attributed to the distinct dynamical response regimes.}
	\label{fig:mp_vary_par}
\end{figure*}

In order to further classify the above-discussed response regimes we invoke the mean position of the impurity $\expval{\hat{x}^I(t)}$, see Figure \ref{fig:vNE_mp_cmp}(a), for different impurity-medium interactions corresponding to the aforementioned dynamical regimes. Evidently, $\expval{\hat{x}^I(t)}$ exhibits individual characteristics in each regime allowing for their clear distinction. For instance, in the case of weak impurity-bath couplings, e.g. $g_{BI}/\tilde{E}\tilde{x}=-0.2$, an oscillatory behavior of $\expval{\hat{x}^I(t)}$ takes place as expected from $\rho^{(1)}_I(x,t)$ [cf. Figure \ref{fig:gpop}(c1)]. Turning to the total reflection regime, e.g. for $g_{BI}/\tilde{E}\tilde{x}=2.0$, $\expval{\hat{x}^I(t)}$ captures the irregular behavior of the impurity on the left edge of the double-well, thus indicating its total reflection from the bosonic environment.
The long-time evolution of $\expval{\hat{x}^I(t)}$ for $g_{BI}/\tilde{E}\tilde{x}=-0.2$ is illustrated in the inset of Figure \ref{fig:vNE_mp_cmp}(a). As can be seen, the decreasing amplitude of $\expval{\hat{x}^I(t)}$ becomes evident which is caused by the continuous energy transfer from the impurity to the bosonic medium, see also Appendix \ref{ap:energy}.

To provide the complete response phase diagram of the impurity we show the behavior of $\expval{\hat{x}^I(t)}$ in dependence of the impurity-medium interaction strength $g_{BI}$, thus capturing the dynamical crossover between the aforementioned regimes, see Figure \ref{fig:vNE_mp_cmp}(b). For convenience the five identified response regimes are labeled from I to V. Regime I corresponds to the steady bound state formation, see the small amplitude oscillations of $\rho^{(1)}_I(x,t)$ in the vicinity of the left site of the double-well. In regime II we find the expected behavior of $\expval{\hat{x}^I(t)}$ represented by its low frequency oscillations around the trap center as shown in Figure \ref{fig:vNE_mp_cmp}(a). For weak impurity-medium interaction strengths, corresponding to regime III, the dissipative oscillatory motion of $\expval{\hat{x}^I(t)}$ occurs characterized by a relatively large amplitude of the underlying oscillations. Increasing $g_{BI}$ to intermediate repulsive values we reach the pinning regime (cf. regime III) where the mean position saturates towards $x^I=0$. For larger $g_{BI}$ the impurity is not able to penetrate the bath anymore and it is totally reflected at the edge of the latter. This regime corresponds to the total reflection one and it is denoted by V in Figure \ref{fig:vNE_mp_cmp}(b).
Finally, we comment on the response regimes in which the mean position obtains values close to zero, viz. the pinning and the tunneling regime. Even though the impurity's mean position within these regimes is well distinguishable [see Figure \ref{fig:vNE_mp_cmp}(a)] in a corresponding experiment a clear distinction might be challenging. To ensure a clear distinction between these regimes one can use the experimentally accessible position variance $s=\expval{\hat{x}}^2-\expval{\hat{x}^2}$ \cite{catani2012, ronzheimer2013}. Since in the tunneling regime the impurity is distributed over the double-well the respective variance is larger than the one in the pinning regime where the impurity is localized at the trap center (not shown here).

In order to expose the robustness of the impurity dynamics with respect to parametric variations we present in Figure \ref{fig:mp_vary_par} $\expval{\hat{x}^I(t)}$ for a wide range of system parameters. As we emphasized previously, it is possible to distinguish between the dynamical response regimes by inspecting the behavior of $\expval{\hat{x}^I(t)}$. Therefore, we choose the impurity-bath coupling strength $g_{BI}$ such that we obtain a behavior of $\expval{\hat{x}^I(t)}$ which can be in turn associated with a specific dynamical response regime.
Figures \ref{fig:mp_vary_par}(a) and (b) show $\expval{\hat{x}^I(t)}$ obtained for $x_0^I/\tilde{x}=5$ and $x_0^I/\tilde{x}=10$, respectively. Here, each mean position exhibits the same behavior as the corresponding one depicted in Figure \ref{fig:vNE_mp_cmp}(a) for $x_0^I/\tilde{x}=8$. As expected within the dissipative oscillation regime, an amplification of the oscillation amplitude occurs as the initial displacement increases. Based on these observations we conclude that the impurity dynamics is robust with respect to the initial displacement and, more precisely for values from $x_0^I=5$ to 10. We remark that in the limit of small displacements $x_0^I$ and intermediate repulsive interaction strengths the ground state is altered, viz. the impurity is initially located between the two one-body density maxima of the bath where it remains in the course of the evolution.

Varying the intraspecies coupling strength $g_{BB}$ between the bath particles we are again able to realize the respective dynamical response regimes but for shifted $g_{BI}$ [see Figures \ref{fig:mp_vary_par}(c) and (d)]. For smaller intraspecies interaction strengths, e.g. $g_{BB}/\tilde{E}\tilde{x}=0.2$, the dynamical regimes are shifted towards smaller absolute values of $g_{BI}$ [cf. Figure \ref{fig:mp_vary_par}(c)] and vice versa in case of a larger $g_{BB}$, e.g. $g_{BB}/\tilde{E}\tilde{x}=1.0$ [cf. Figure \ref{fig:mp_vary_par}(d)]. In particular, in order to realize the steady bound state regime for a larger $g_{BB}$ stronger impurity-medium attractions are necessary than in the case of a weakly interacting medium (small $g_{BB}$). We attribute this property to the mobility of the bath particles, i.e. the compressibility of the medium, which becomes smaller (larger) for increasing (decreasing) $g_{BB}$. Therefore, in the case of a strongly interacting bath larger impurity-medium attractions are needed in order to shift a sufficient amount of the medium's one-body density to the left site of the double-well which, eventually, binds the impurity [see also the mean positions corresponding to the steady bound state regime in Figures \ref{fig:mp_vary_par}(c) and (d)]. On the other hand, for strongly repulsive impurity-medium interactions the total reflection regime emerges in the case of a weakly interacting bath at smaller $g_{BI}$ compared to the case of a strongly interacting bath. We attribute this property to the $g_{BB}$-dependence of the spatial extension of the medium's cloud. The latter is broadened for large $g_{BB}$ and becomes narrower at the sites of the double-well for small $g_{BB}$ leading in the latter case to an increased effective potential barrier experienced by the impurity [see also Eq. (\ref{eq:TAEP})]. Therefore, it is easier for the impurity to overcome the bosonic medium at the left site of the double-well in the case of a larger $g_{BB}$, i.e. for a broadened background, than in the case of a smaller $g_{BB}$, e.g. compare the mean positions of $g_{BI}/\tilde{E}\tilde{x}=2.0$ in Figures \ref{fig:mp_vary_par}(c) and (d).

Hence, the intraspecies interaction strength $g_{BB}$ indeed impacts the impurity dynamics. However, the same dynamical regimes can be recaptured by properly adjusting $g_{BI}$ at least in the considered cases of relatively weak and strong intraspecies interaction strengths, i.e., $g_{BB}/\tilde{E}\tilde{x} = 0.2, 1.0$ considered herein. We remark that for even stronger repulsions where the medium resides in a Mott-like state an altered dynamical response of the impurity is expected, an investigation which is left for future studies.

Subsequently, we aim to quantify the associated impurity-medium entanglement by monitoring the von-Neumann entropy \cite{paskauskas2001}, which reads
\begin{equation}
S^{VN}(t)= -\sum_{i=1}^M \lambda_i(t) \ln \lambda_i(t).
\end{equation}
This expression possesses an upper bound for maximal entanglement between the species, viz. $\lambda_i=1/M$ leading to $S_{\textrm{max}}^{VN} = \ln M=1.79$ in our case ($M=6$), and vanishes when no entanglement is present, e.g. $\lambda_i=1$ with $\lambda_{i>1}=0$.
In Figure \ref{fig:vNE_mp_cmp}(c) we provide the long-time evolution of the von-Neumann entropy for different values of $g_{BI}$ corresponding to the five dynamical response regimes of the impurity. We find in all regimes a finite impurity-medium entanglement \cite{mistakidis2019, mistakidis2019c} which tends to saturate for larger times ($t/\tilde{\omega}^{-1}>1500$) besides the tunneling regime where $S^{VN}(t)$ performs an oscillatory motion.
Among the investigated regimes the steady bound state and the total reflection regime appearing at large attractive and repulsive $g_{BI}$ couplings experience the smallest amount of entanglement. Indeed, $S^{VN}(t)$ is maximized within the pinning ($g_{BI}/\tilde{E}\tilde{x}=1.0$) and the dissipative oscillation regime ($g_{BI}/\tilde{E}\tilde{x}=-0.2$). Additionally, in the latter response regime the entanglement increases with time and reaches a plateau at around $t/\tilde{\omega}^{-1}=300$. During this time-interval the impurity penetrates the bosonic medium twenty times, thereby, enhancing the entanglement at each penetration. In the pinning regime, the system becomes maximally entangled after the impurity penetrates its environment a single time and, subsequently, becomes pinned between the effective barriers raised by the bosonic medium.

To obtain a better understanding of the underlying microscopic mechanisms appearing in the respective response regimes we analyze, in the following, the impurity dynamics with respect to an effective potential \cite{mistakidis2019, mistakidis2020, theel2020}, which reads
\begin{align}
V_{I}^{\textrm{eff}}(x^I) = \frac{1}{T} \int_0^T \left[ N_B g_{BI} \rho_B^{(1)}(x^I, t) + V_I(x^I)\right] \textrm{d}t.
\label{eq:TAEP}
\end{align}
Here, we choose $x_0^I=0$ for the harmonic confinement $V_I(x^I)$. This effective potential is based on the assumption of a product state ansatz $\ket{\Psi^{\textrm{MB}}(t)} = \ket{\Psi^B(t)}\otimes \ket{\Psi^I(t)}$ where the degrees of freedom of the bosonic medium are integrated out. Thus, the effective potential is the superposition of the harmonic confinement $V_I(x^I)$ after the quench and the one-body density of the environment weighted with the particle number $N_B$ and the impurity-medium interaction strength $g_{BI}$ \cite{theel2020}. Since the bosonic medium remains to a certain extend well localized at the sites of the double-well in the course of the evolution the averaging of the effective potential over the total propagation time $T/\tilde{\omega}^{-1}=200$ is justified. Note that even though we considered a product state ansatz for the construction of $V_{I}^{\textrm{eff}}(x^I)$, beyond mean-field effects are included in the one-body density $\rho_B^{(1)}(x,t)$ which is calculated with a many-body ansatz (see section \ref{sec:ML-X}).

The time-averaged effective potentials (gray solid lines) for four representative dynamical response regimes corresponding to $g_{BI}/\tilde{E}\tilde{x} = -0.9,-0.7,1.0,2.0$ are demonstrated in Figures \ref{fig:TAEP_sOccup}(a1)-(d1).
For the steady bound state regime, at $g_{BI}/\tilde{E}\tilde{x} = -0.9$ the effective potential $V_{I}^{\textrm{eff}}(x^I)$ takes the form of an asymmetric double-well [cf. Figure \ref{fig:TAEP_sOccup}(a1)] with a deeper left site since $\rho_B^{(1)}(x,t)$ is attracted to the impurity and, therefore, it is shifted to the left site of the double-well.
For intermediate attractions, e.g. $g_{BI}/\tilde{E}\tilde{x} = -0.7$, $V_{I}^{\textrm{eff}}(x^I)$ has the shape of a nearly symmetric double-well potential [see Figure \ref{fig:TAEP_sOccup}(b1)]. In this case the symmetry can be ascribed to the fact that the averaging in Eq. (\ref{eq:TAEP}) is performed over the period $T/\tilde{\omega}^{-1}=200$ during which the maximum of $\rho_B^{(1)}(x,t)$ shifts from one site of the double-well to the other leading on average to the observed nearly symmetric double-well. This in turn explains the tunneling behavior of the impurity depicted in Figure \ref{fig:gpop}(b1).

This picture changes drastically for repulsive impurity-medium interactions. Here, the bosonic medium imprints a potential barrier with two maxima located at the two sites of its actual double-well.
Therefore, in this case, e.g. for $g_{BI}/\tilde{E}\tilde{x} = 1.0$, the effective potential $V_{I}^{\textrm{eff}}(x^I)$ obtains the shape of a deformed harmonic oscillator having an additional prominent dip at the trap center [Figure \ref{fig:TAEP_sOccup}(c1)].
As $g_{BI}$ increases to $g_{BI}/\tilde{E}\tilde{x} = 2.0$ the aforementioned two density maxima of $\rho_B^{(1)}(x,t)$ become visible giving rise to two potential barriers. Due to the superposition of the latter with the initially considered harmonic confinement $V_I(x^I)$ the effective potential of the impurity $V_{I}^{\textrm{eff}}(x^I)$ deforms to an asymmetric triple-well [see Figure \ref{fig:TAEP_sOccup}(d1)].

As a next step, we construct for each $V_{I}^{\textrm{eff}}(x^I)$ the single-particle Hamiltonian $\hat{\mathcal{H}}^{(1)} = \frac{\hbar^2}{2m_I} \frac{\partial^2}{\partial x^2} + \hat{V}_{I}^{\textrm{eff}}$ and calculate its first six energetically lowest-lying eigenfunctions $\psi_i^{\textrm{eff}}$. The corresponding absolute squares of the eigenfunctions shifted by their eigenenergies are shown in Figures \ref{fig:TAEP_sOccup}(a1)-(d1).
In the following, theses sets of eigenfunctions $\psi_i^{\textrm{eff}}$ are taken as basis sets in order to analyze the underlying microscopic mechanisms in the course of the impurity dynamics. The time-dependent probability for the impurity to occupy the state $\psi_i^{\textrm{eff}}$ reads
\begin{align}
P_i^{1}(t) = \sum_{j} \rvert\braket{\Psi^{\textrm{MB}}(t)}{\varphi_j^B}\otimes\ket{\psi_i^{\textrm{eff}}}\lvert^2,
\end{align}
where $\{\varphi_j^B\}$ is an arbitrary basis set covering the whole subspace of the bosonic medium. By summing over all basis states of the bath we single out the probability to find the impurity in $\psi_i^{\textrm{eff}}$. Note, that $\ket{\Psi^{\textrm{MB}}(t)}$ is the full many-body wave function defined via ML-MCTDHX [Eq. (\ref{eq:schmidt_decomp})], while $\psi_i^{\textrm{eff}}$ serves as a basis set to unravel the underlying participating dynamical processes. The occupation probabilities for the respective basis sets [Figures \ref{fig:TAEP_sOccup}(a1)-(d1)] are presented in Figures \ref{fig:TAEP_sOccup}(a2)-(d2). In order to justify the quality of the basis we sum up all non-vanishing occupation probabilities $P_i^{1}(t)$ and determine the time at which the sum exceeds and subsequently remains above 0.94 (dashed gray lines).

For $g_{BI}/\tilde{E}\tilde{x} = -0.9$ the impurity predominantly populates the energetically lowest eigenstate on the left site of the tilted effective double-well $V_{I}^{\textrm{eff}}(x^I)$, while the probability to occupy energetically higher-lying states is strongly suppressed as time evolves [see Figure \ref{fig:TAEP_sOccup}(a2)]. Based on this behavior of $P_i^{1}(t)$, i.e. the spatial localization of the impurity, and the fact that the eigenenergies are negative we associate the bound state formation with the energetically lowest eigenstate of the effective potential $V_{I}^{\textrm{eff}}(x^I)$ \cite{mistakidis2019}. A further analysis of this steady bound state is provided in Appendix \ref{ap:steady_bound_state_regime} in terms of the involved two-body density.
For intermediate attractive impurity-medium couplings corresponding to the tunneling regime $V_{I}^{\textrm{eff}}(x^I)$ has, in contrast to the steady bound state regime, the shape of a nearly symmetric double-well. Accordingly, the impurity dynamics is mainly determined by the superposition of the two energetically lowest eigenstates of $V_{I}^{\textrm{eff}}(x^I)$ [see Figure \ref{fig:TAEP_sOccup}(b2)].

In the case of intermediate repulsive interactions a pinning of the impurity between the density maxima of the bosonic bath is realized [cf. Figure \ref{fig:TAEP_sOccup}(c2) with $g_{BI}/\tilde{E}\tilde{x} = 1.0$]. Here, the occupation probabilities start to saturate after $t/\tilde{\omega}^{-1}=50$ which in turn leads to the energetically lowest eigenstate of $V_{I}^{\textrm{eff}}(x^I)$ being predominantly populated. However, we observe that the probability to find the impurity in an energetically higher-lying eigenstate is approximately $40\,\%$. In this sense, the broadening of $\rho_I^{(1)}(x,t)$ around the trap center [see Figure \ref{fig:gpop}(e1)] can be interpreted as impurity excitations with respect to the effective potential. We refer to those excitations as \textit{hidden excitations} since they can only be identified by such a microscopic analysis. It is worth noticing that the spatial structure of the first three species functions of the impurity $\ket{\Psi_i^I}$ are in a good agreement with the three energetically lowest eigenfunctions of $V_{I}^{\textrm{eff}}(x^I)$ (see Appendix \ref{ap:pinning_regime}).
For strong repulsive couplings, e.g. $g_{BI}/\tilde{E}\tilde{x} = 2.0$, the effective potential exhibits the shape of a triple-well where the two potential barriers stem from the bosonic medium being localized at the sites of the double-well potential. Since these potential barriers are comparatively large, the impurity is totally reflected by the left barrier and predominantly occupies the eigenstates located in the left site of the triple-well [see Figure \ref{fig:TAEP_sOccup}(d2)].

In summary, the analysis of the effective potential enables us to unravel the underlying microscopic mechanisms of the impurity dynamics. In particular, it allows for a deep understanding of the steady bound state formation and proves to be crucial in order to identify the hidden excitations in the pinning regime.
Note that for an analogous analysis of the dissipative oscillation regime a much larger set of eigenfunctions has to be taken into account to achieve a comparable quality of the employed single-particle basis (similar to a coherent state in a harmonic oscillator). In Appendix \ref{ap:diss_osc_reg} we provide a discussion of the impurity dynamics in the dissipative oscillation regime where we compare the motion of its mean position to a damped harmonic oscillator.

\section{Dynamical response regimes of two impurities}
\label{sec:dyn_reg_two_imp}

To generalize our findings, in the following, we consider two non-interacting impurities ($g_{II}=0$) coupled to the bosonic environment. Therefore, all interactions between the impurities are induced by their coupling to the bosonic medium \cite{chen2018, dehkharghani2018, mistakidis2020}. The quench protocol is the same as described in section \ref{sec:setup}.

\begin{figure}[t]
	\centering
	\includegraphics[width=1\linewidth]{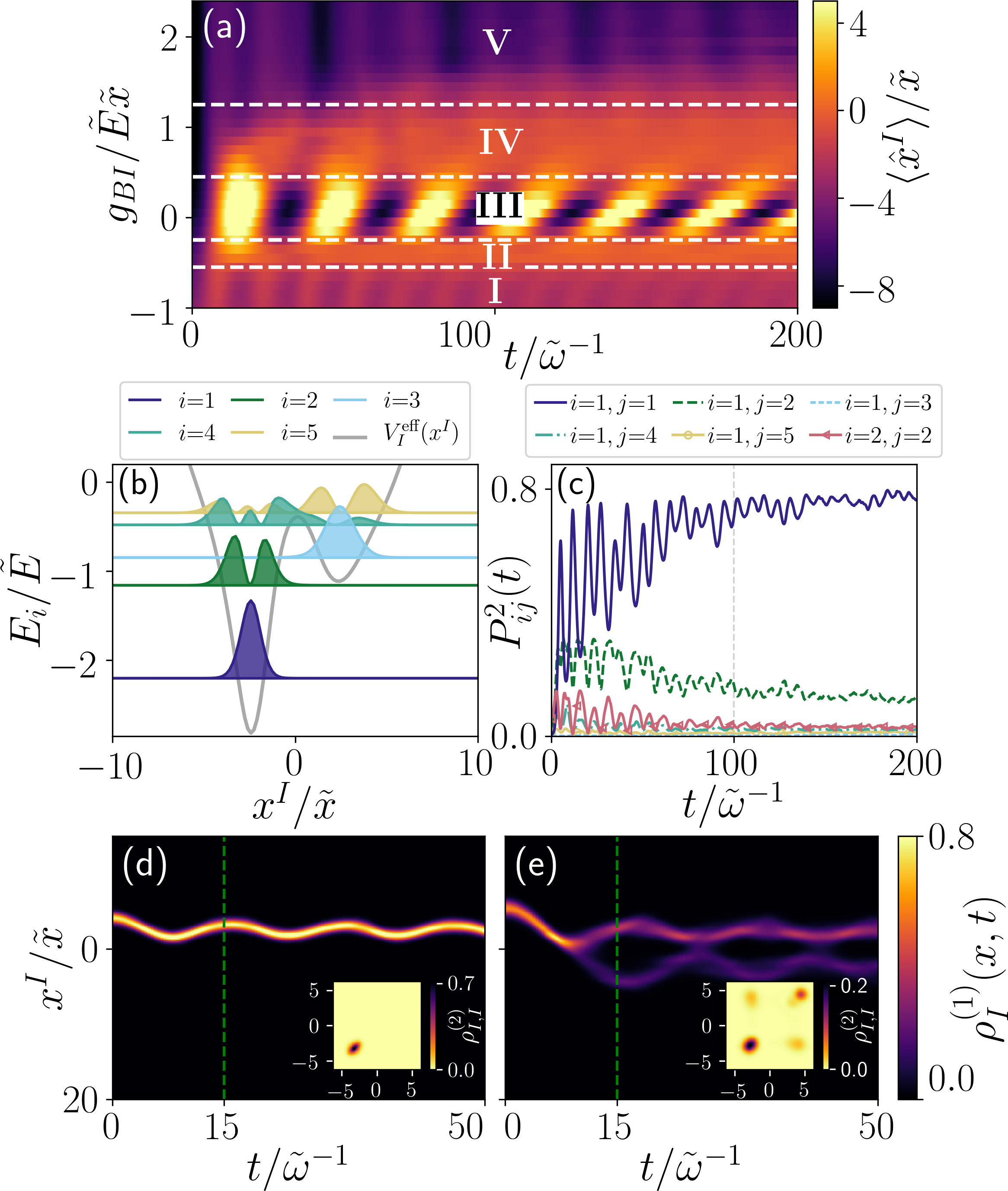}
	\caption{(a) Temporal evolution of the two impurities' mean position for various impurity-medium coupling strengths $g_{BI}$. (b) Time-averaged effective potential calculated for the two-impurity case for $g_{BI}/\tilde{E}\tilde{x} = -0.9$ [see Eq. (\ref{eq:TAEP})]. The associated eigenfunctions $\psi_i^{\textrm{eff}}$ of the effective potential are shifted with respect to their eigenenergies $E_i$. (c) Time-evolution of the conditional probability $P_{ij}^2(t)$ to find one impurity in $\psi_i^{\textrm{eff}}$ while at the same time the other impurity occupies the eigenstate $\psi_j^{\textrm{eff}}$. Panels (d) and (e) showcase the time-evolution of the one-body densities $\rho_{I}^{(1)}(x,t)$ for $g_{BI}/\tilde{E}\tilde{x} = -1.0$ and $g_{BB}/\tilde{E}\tilde{x} = 0.5,1.0$, respectively. Each inset illustrates a snapshot of the respective two-body density $\rho_{I,I}^{(2)}(x_1^I,x_2^I)$ at $t/\tilde{\omega}^{-1}=15$ (green dashed line).}
	\label{fig:two_impurities}
\end{figure}

In order to characterize the dynamics of the two bosonic impurities we monitor the time-evolution of their mean position, i.e. their center of mass position, for different impurity-bath coupling strengths $g_{BI}$ [Figure \ref{fig:two_impurities}(a)]. Analogously to the one-impurity case we identify five dynamical response regimes depending on $g_{BI}$ and appearing for similar interaction strengths as for $N_I=1$, see also Figure \ref{fig:vNE_mp_cmp}(b).
However, we find in the response regime II [Figure \ref{fig:two_impurities}(a)] that the mean position is almost constant with $\langle x^I \rangle = 0$, whereas in the single-impurity case $\expval{\hat{x}^I(t)}$ oscillates with a small amplitude around zero. By inspecting the impurity's one-body density $\rho_I^{(1)}(x,t)$ for the corresponding regime II (not shown here) it is observed that $\rho_I^{(1)}(x,t)$ is almost equally distributed over the effective double-well potential (mediated by the bosonic bath) yielding a mean position close to zero. Moreover, the discrepancy between the tunneling and the pinning regime regarding the impurities' mean position can be revoked by inspecting the variance $s$ analogously to the single-impurity case [see also Figure \ref{fig:vNE_mp_cmp}(b)]. Additionally, we observe a broadening of the transition from the pinning to the total reflection regime for increasing $g_{BI}$ with respect to the single-impurity case where we identified a sharper transition [see Figure \ref{fig:two_impurities}(a)]. As a case example we shall focus on the steady bound state regime and unravel the microscopic processes in the case of two impurities.

To provide a qualitative understanding of the dynamics in the steady bound state response regime we will describe the impurity dynamics within an effective potential picture, similarly to the previously discussed single-impurity case.
For this purpose, we calculate the effective potential $V_{I}^{\textrm{eff}}(x^I)$ [see Eq. (\ref{eq:TAEP})] for $g_{BI}/\tilde{E}\tilde{x} = -0.9$ which we present in Figure \ref{fig:two_impurities}(b) \footnote{Note, that Eq. (\ref{eq:TAEP}) can be also employed in the case of two impurities.}. Additionally, we compute the effective single-particle eigenstates $\psi_i^{\textrm{eff}}$ shifted by their eigenenergies $E_i$ [see Figure \ref{fig:two_impurities}(b)].
A comparison with the effective potential in Figure \ref{fig:TAEP_sOccup}(a1) reveals that the left site of $V_{I}^{\textrm{eff}}(x^I)$ in the two-impurity case is much deeper. Therefore, the two energetically lowest eigenfunctions $\psi_1^{\textrm{eff}}$ and $\psi_2^{\textrm{eff}}$ in the two-impurity scenario are located at the left site of $V_{I}^{\textrm{eff}}(x^I)$.

Next, we unravel the interplay between the impurities by studying the conditional probabilities to occupy specific eigenstates of their effective potential. In particular, we define $P_{ij}^2(t)$ as the probability for one impurity to occupy the effective eigenstate $\psi_i^{\textrm{eff}}$ while at the same time the other impurity populates the eigenstate $\psi_j^{\textrm{eff}}$.

In Figure \ref{fig:two_impurities}(c) we demonstrate $P_{ij}^2(t)$ with respect to the five energetically lowest-lying eigenstates of the effective potential for $g_{BI}/\tilde{E}\tilde{x} = -0.9$ [cf. Figure \ref{fig:two_impurities}(b)] \footnote{The sum of the presented probabilities exceeds 0.94 at $t/\tilde{\omega}^{-1} \approx 100$ (dashed gray line).}.
Interestingly, we can infer that the two impurities predominantly occupy simultaneously the same energetically lowest eigenstate, which is in accordance with the observations made in the single-impurity case [Figure \ref{fig:TAEP_sOccup}(a2)]. However, also single-particle excitations in the second eigenstate $\psi_2^{\textrm{eff}}$ [$P_{12}^2(t)$] as well as two particle excitations contribute to the many-body wave function of the impurities \cite{keiler2019, theel2020}. For instance, we find a small but non-vanishing probability for observing two impurities in the second eigenstate $P_{22}^2(t)$.
Notice that an analogous analysis for the other response regimes, apart from the dissipative oscillation one, leads to similar observations where one or two impurities occupy simultaneously the same or different excited eigenstates of $V_{I}^{\textrm{eff}}(x^I)$ (not shown here).

In the following, we demonstrate that the dynamical response of two impurities can be tuned by changing the intraspecies interactions $g_{BB}$ of the bath, similarly to the single-impurity case (cf. Section \ref{sec:dyn_reg_single_imp}). As a characteristic example we present in Figures \ref{fig:two_impurities}(d) and (e) the time-evolution of the impurities' one-body densities for strong impurity-medium attractions and two different intraspecies couplings $g_{BB}$. In the case of a weakly interacting bath [cf. Figure \ref{fig:two_impurities}(d)] the major portion of the medium's one-body density is shifted to the left site of the double-well (not shown here) such that the impurities are permanently bound to the medium at this site. Therefore, this situation corresponds to the steady bound state regime [cf. regime I in Figure \ref{fig:two_impurities}(a)]. By increasing the intraspecies interaction strength the compressibility of the bath is reduced and, thus, also the amount of the medium's one-body density accumulated at the left double-well site. In this sense, the impurities cannot be permanently bound at one site of the double-well and distribute over the latter performing an effective tunneling dynamics \footnote{In order to realize the dynamics corresponding to the steady bound state regime occurring for large $g_{BB}$ the impurity-medium attraction needs to be adjusted accordingly. For instance, we have verified for $g_{BB}/\tilde{E}\tilde{x}=1.0$ and $g_{BI}/\tilde{E}\tilde{x}=-2.0$ the formation of a steady bound state response.}. In particular, the corresponding one-body density $\rho_{I}^{(1)}(x,t)$ splits into two branches with each one oscillating at an individual site of the double-well [see Figure \ref{fig:two_impurities}(e)]. Thereby, this dynamical response of the impurities resembles the dynamics corresponding to regime II in Figure \ref{fig:two_impurities}(a) in the case of a weakly interacting medium.

To further shed light on the spatial configuration of the two impurities we investigate their reduced two-body density $\rho_{I,I}^{(2)}(x_1^I,x_2^I)$, defined as
\begin{align}
&\rho_{I,I}^{(2)}(x_1^I,x_1^I)\nonumber \\
&= \int \textrm{d}x_1^B\cdots\textrm{d}x_{N_B}^B \left| \Psi^{\textrm{MB}}(x_1^I,x_2^I,x_2^B,\dots,x_{N_B}^B) \right|^2.
\end{align}
In the inset of Figure \ref{fig:two_impurities}(d) the reduced two-body density of the impurities is provided for a weakly interacting bath at a specific time instant $t/\tilde{\omega}^{-1}=15$. It exhibits an elongated peak along the diagonal which is, in particular, located at the left double-well site. This behavior indicates non-vanishing impurity-impurity induced correlations \cite{mistakidis2020}. On the other hand, for a strongly interacting medium and at $t/\tilde{\omega}^{-1}=15$ (corresponding to the time instance where the one-body density features the splitting) the two-body density reveals two dominant maxima in the diagonal and in terms of the amplitude two smaller ones in the off-diagonal [see inset of Figure \ref{fig:two_impurities}(e)]. Thereby, the diagonal peaks of $\rho_{I,I}^{(2)}(x_1^I,x_2^I)$ explicate that both impurities move together and reside in either of the double-well sites. Complementary, the smaller off-diagonal peaks hint at a suppressed probability of finding each impurity at a different site. Thus, a scenario in which one impurity remains at one double-well site while the other impurity tunnels to the other one is less probable.

In summary, we have deduced that the dynamical response regimes of two non-interacting impurities are similar to the single-impurity case with a small modification regarding the effective tunneling regime. Furthermore, we can gain insights into the time-dependent microscopic configuration of the two impurities by investigating the associated conditional probability to find the impurities in two particular eigenstates of their effective potential. Particularly, we exemplified that for the steady bound state regime the impurities predominantly occupy simultaneously the lowest-lying eigenstate of $V_{I}^{\textrm{eff}}(x^I)$. However, we also observed the occurrence of single- and two-particle excitations in higher-lying eigenstates. Additionally, for strong impurity-medium attractions we varied the compressibility of the bath by considering a weakly and a strongly interacting medium. Thereby, we observed an alteration of the dynamical response of the impurities from a steady bound state (for small $g_{BB}$) to a distribution of the impurities over the double-well (for large $g_{BB}$). The respective two-body density distributions revealed that in the latter case the impurities tend to move together manifesting the dominant presence of attractive induced interactions during the effective tunneling dynamics of the impurities.

\section{Summary and Outlook}
\label{sec:summary_outlook}

We have investigated the dynamical behavior of bosonic impurities colliding with a BEC trapped in a double-well. The impurities are initially confined in a harmonic oscillator which is spatially displaced with respect to the double-well of the bosonic medium. Upon quenching the harmonic potential to the trap center of the double-well the quantum dynamics is induced such that the impurities collide with the bosonic environment. The correlated non-equilibrium dynamics is tracked with the variational ML-MCTDHX method which enables us to access the full many-body wave function of the system, thereby, including all relevant inter- and intraspecies correlations.

By varying the impurity-medium interaction strength $g_{BI}$ from strongly attractive to repulsive values we are able to control the collisional dynamics of the impurity and identify five distinct dynamical response regimes by inspecting the associated one-body density evolution. These response regimes correspond to the \textit{steady bound state regime}, the \textit{tunneling regime}, the \textit{dissipative oscillation motion}, the \textit{pinning regime} and the \textit{total reflection regime}.
We demonstrate that they can be easily identified by monitoring the mean position of the impurity. Moreover, by calculating a cross-over phase diagram of the impurity's mean position with respect to the impurity-medium coupling strength we obtain an overview of the emergent dynamical response regimes as a function of $g_{BI}$ and identify smooth transitions between two consecutive ones. Additionally, we explicate the robustness of the response regimes for different parametric system variations, i.e. the intraspecies interaction strength of the bath and the initial displacement of the impurity's harmonic trap.

To provide a better understanding of the involved microscopic mechanisms we employ a time-averaged effective potential picture. By projecting the total many-body wave function onto the eigenstates associated with this effective potential allows us to gain insights into the underlying excitation processes for different interactions. In particular, we find that the impurity is bound to the bosonic medium for strong attractive impurity-bath interaction strengths corresponding to the steady bound state regime and unveil hidden excitations in the pinning regime occurring for intermediate repulsive $g_{BI}$.
We extend our study to the two-impurity case where we showcase the emergence of similar dynamical response regimes as in the single-impurity scenario. Furthermore, we unravel the underlying microscopic mechanisms of the impurities' dynamics analogously to the single-impurity case. Here, for the steady bound state regime the participation of single- as well as two-particle excitations into energetically higher-lying states of the effective potential is demonstrated. Additionally, for strong impurity-medium attractions we show that the dynamical response of the impurities can be altered from a steady bound state (for a weakly interacting medium) to a configuration where the impurities distribute over the double-well (for a strongly interacting medium).

The results of this work are beneficial for future ultracold atom experiments of impurity-medium scattering for investigating the corresponding collisional channels caused exclusively by presence of the impurity-bath entanglement. Furthermore, this setup can be extended by implementing an additional spin-degree-of-freedom for two non-interacting or weakly-interacting impurities. In this case it would be interesting to identify the individual spin configurations and related spin-mixing processes in dependence of the impurity-medium coupling and whether the impurities evolve as a `Cooper-pair'. Certainly the generalization of our results to higher dimensions is an intriguing perspective.

\appendix
\section{Impurity-medium energy transfer processes}
\label{ap:energy}

Due to the initial quench of the harmonic oscillator potential the impurities collide with their bosonic background and, thereby, a transfer of energy to the latter is triggered \cite{mistakidis2019c, mukherjee2020, mistakidis2020a}. The total energy of the system can be written as $E_{\textrm{tot}} = E_{\textrm{tot}}^B + E_{\textrm{tot}}^I + E_{\textrm{int}}$ where $E_{\textrm{tot}}^\sigma=\bra{\Psi^{\textrm{MB}}}\mathcal{\hat{H}}^\sigma \ket{\Psi^{\textrm{MB}}}$ represents the total energy of species $\sigma\in\{B,I\}$ and $E_{\textrm{int}}=\bra{\Psi^{\textrm{MB}}}\mathcal{\hat{H}}^{\textrm{int}} \ket{\Psi^{\textrm{MB}}}$ the interaction energy between the species.

\begin{figure}[t]
	\centering
	\includegraphics[width=.95\linewidth]{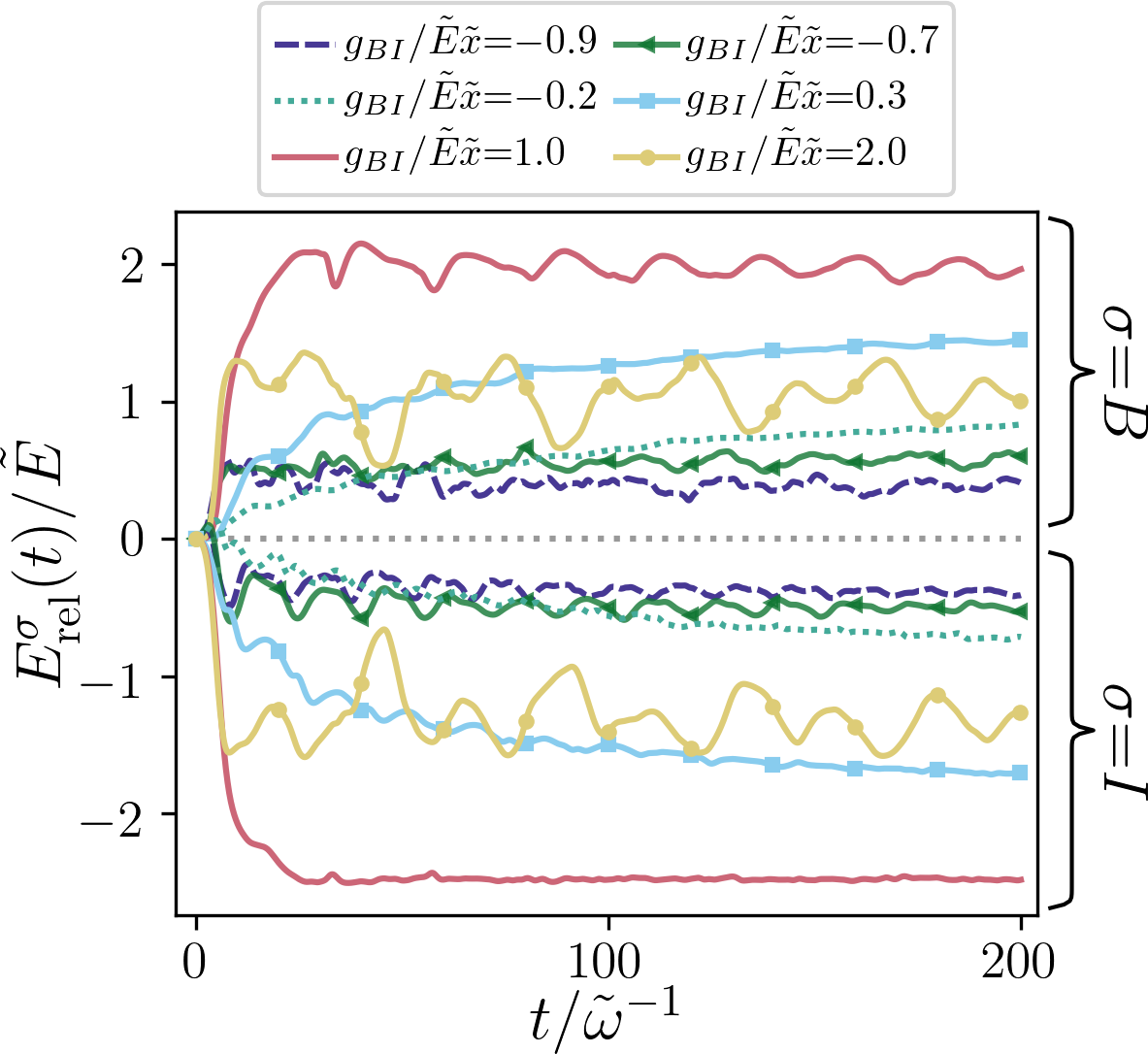}
	\caption{Time-evolution of the relative energy $E_{\textrm{rel}}^\sigma(t) = E_{\textrm{tot}}^\sigma(t) - E_{\textrm{tot}}^\sigma(0)$ of species $\sigma\in\{B,I\}$ for different $g_{BI}$ (see legend). In each dynamical response regime the impurity transfers energy to the bosonic medium.}
	\label{fig:ap_energy}
\end{figure}

In order to capture the quench-induced impurity-medium energy transfer we present in Figure \ref{fig:ap_energy} the relative energy of species $\sigma$ defined as $E_{\textrm{rel}}^\sigma(t) = E_{\textrm{tot}}^\sigma(t) - E_{\textrm{tot}}^\sigma(0)$. In each dynamical response regime we observe an energy transfer from the impurity to its environment. The smallest energy transfer occurs for attractive impurity-medium interaction strengths where, due to the attraction to the bath, the impurity resides closer to the trap center leading to a smaller initial total energy $E_{\textrm{tot}}^I(0)$ than in the case of repulsive $g_{BI}$ [compare with the mean position in Figure \ref{fig:vNE_mp_cmp}(a)].

For weak impurity-bath coupling strengths of either sign, i.e. $g_{BI}/\tilde{E}\tilde{x}=-0.2,0.3$, the impurity continuously transmits energy to its environment until the relative energy of both species eventually saturates for longer times \cite{mukherjee2020, mistakidis2020a}. This loss of impurity energy essentially causes its dissipative oscillatory behavior [cf. Figure \ref{fig:vNE_mp_cmp}(a)].
The largest energy transfer takes place in the pinning regime. Here, the impurity overcomes the bosonic medium in the left site of the double-well only once and, thereby, transfers a large amount of energy to the bosonic medium such that the impurity becomes pinned within the latter.

\section{Two-body correlation in the steady bound state regime}
\label{ap:steady_bound_state_regime}

To elucidate the interplay between the bosonic medium and the impurity in the steady bound state regime in more detail we perform an analysis of the impurity-medium reduced two-body density $\rho_{B,I}^{(2)}(x^I,x^B;t)$, defined as
\begin{align}
&\rho_{B,I}^{(2)}(x^I,x^B;t)\nonumber \\
&= \int \textrm{d}x_2^B\cdots\textrm{d}x_{N_B}^B \left| \Psi^{\textrm{MB}}(x^I,x^B,x_2^B,\dots,x_{N_B}^B;t) \right|^2.
\end{align}
This quantity provides information about the probability of finding the impurity at position $x^I$ and one particle of the bosonic background located at $x^B$.

\begin{figure}[t]
	\centering
	\includegraphics[width=1\linewidth]{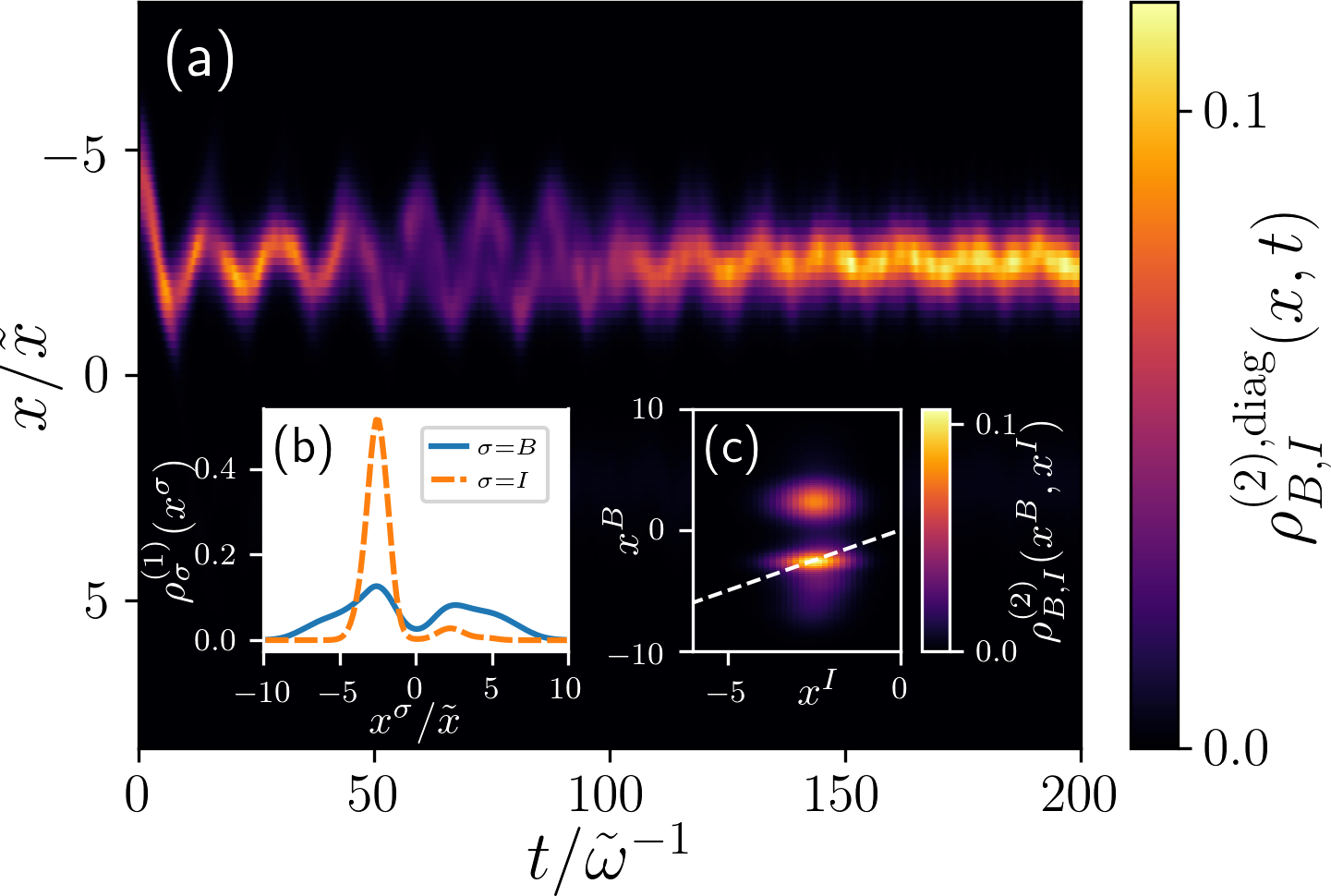}
	\caption{(a) Time-evolution of the diagonal elements of the two-body density $\rho_{B,I}^{(2),\textrm{diag}}(x,t)$ for $g_{BI}/\tilde{E}\tilde{x}=-0.9$ corresponding to the steady bound state regime. (b)-(c) Snapshot of the one-body density matrix $\rho_{\sigma}^{(1)}(x^\sigma)$ with $\sigma\in\{B,I\}$ and the two-body density $\rho_{B,I}^{(2)}(x^I,x^B)$, respectively, at $t/\tilde{\omega}^{-1}=150$.}
	\label{fig:ap_dmat2}
\end{figure}

A snapshot of the one-body density $\rho_{\sigma}^{(1)}(x^\sigma)$ with $\sigma\in\{B,I\}$ and the two-body density $\rho_{B,I}^{(2)}(x^I,x^B)$, respectively, at $t/\tilde{\omega}^{-1}=150$ and for $g_{BI}/\tilde{E}\tilde{x}=-0.9$ is depicted in Figures \ref{fig:ap_dmat2}(b) and (c). Here, we find the impurity to be localized at the left maximum of the bosonic medium corresponding to the left site of the double-well which agrees with the observations made for the single-impurity case in Figures \ref{fig:TAEP_sOccup}(a1) and (a2). Furthermore, the two-body density indicates that the probability to find one particle of the bath at the left site of the double-well, i.e. close to the impurity, is enhanced compared to the respective probability for the right site. In particular, the diagonal of $\rho_{B,I}^{(2)}(x^I,x^B)$ [dashed white line in Figure \ref{fig:ap_dmat2}(c)] represents the probability to capture the impurity and one particle of the environment at the same position which we will refer to as $\rho_{B,I}^{(2),\textrm{diag}}(x,t)$. Figure \ref{fig:ap_dmat2}(a) shows the time-evolution of $\rho_{B,I}^{(2), \textrm{diag}}(x,t)$ which strongly resembles the one-body density of the impurity for $t/\tilde{\omega}^{-1}>100$ [cf. Figure \ref{fig:gpop}(a1)] and designates a high probability for the impurity and one particle of the bosonic medium to be at the same location \cite{mistakidis2019, mistakidis2020}.

\section{The dissipative oscillation response regime: effective mass and damping of the Bose polaron}
\label{ap:diss_osc_reg}

Let us also analyze the dissipative oscillation regime in the case of a single-impurity in more detail. As observed in Figure \ref{fig:vNE_mp_cmp}(a), the mean position of the impurity for weak impurity-medium couplings exhibits a damped oscillatory behavior. Therefore, in the following we compare the analytical solution of a damped harmonic oscillator with the mean position $\expval{\hat{x}^I(t)}$ and mean momentum $\expval{\hat{p}^I(t)}$ obtained within the ML-MCTDHX method (see section \ref{sec:ML-X}). The equation of motion of a particle subjected to a damped harmonic oscillator \cite{um2002} reads
\begin{equation}
\ddot{x} + \frac{\gamma^{\textrm{eff}}}{m^{\textrm{eff}}} \dot{x} + (\omega^{\textrm{eff}})^2x = 0,
\label{eq:ap_dho_eom}
\end{equation}
where $\gamma^{\textrm{eff}}$ denotes the effective damping constant, $\omega^{\textrm{eff}}$ the effective trapping frequency and $m^{\textrm{eff}}$ refers to the effective mass of the impurity. Here, we interpret the impurity as a quasi-particle, namely a Bose polaron, which is dressed by the excitations of its surrounding and moves in an effective harmonic oscillator. The mean position for a particle obeying Eq. (\ref{eq:ap_dho_eom}) reads
\begin{align}
\expval{\hat{x}_{\textrm{eff}}(t)} &= \exp\!\left(-\frac{\gamma^{\textrm{eff}}}{2m^{\textrm{eff}}}t\right) 
\nonumber \\ \times&
\left[ x_0\cos(\omega_0t) - \frac{x_0\gamma^{\textrm{eff}}}{2\omega_0(m^{\textrm{eff}})^2}\sin(\omega_0t)\right],
\label{eq:ap_dho_mp}
\end{align}
with $\omega_0=\sqrt{(\omega^{\textrm{eff}})^2 - \left(\frac{\gamma^{\textrm{eff}}}{2m^{\textrm{eff}}}\right)^2}$. Additionally, we assume that the particle is initially at rest, i.e. $p_0=0$, and shifted by $x_0^{\textrm{eff}}=\expval{\hat{x}^I(0)}$.
The corresponding mean momentum of Eq. (\ref{eq:ap_dho_eom}) is accordingly written as
\begin{align}
\expval{\hat{p}_{\textrm{eff}}(t)} &= - \exp\!\left(-\frac{\gamma^{\textrm{eff}}}{2m^{\textrm{eff}}}t\right)
\nonumber \\ \times&
\left(m^{\textrm{eff}}\omega_0x_0 + \frac{x_0(\gamma^{\textrm{eff}})^2}{4\omega_0(m^{\textrm{eff}})^2} \right) \sin(\omega_0t).
\label{eq:ap_dho_mm}
\end{align}

Subsequently, we fit the analytical results of Eqs. (\ref{eq:ap_dho_mp}) and (\ref{eq:ap_dho_mm}) to the mean position and mean momentum calculated from the ML-MCTDHX approach for the free parameters $\gamma^{\textrm{eff}}$, $\omega^{\textrm{eff}}$ and $m^{\textrm{eff}}$ \cite{mistakidis2019, mistakidis2019a}.

\begin{figure}[t]
	\centering
	\includegraphics[width=.95\linewidth]{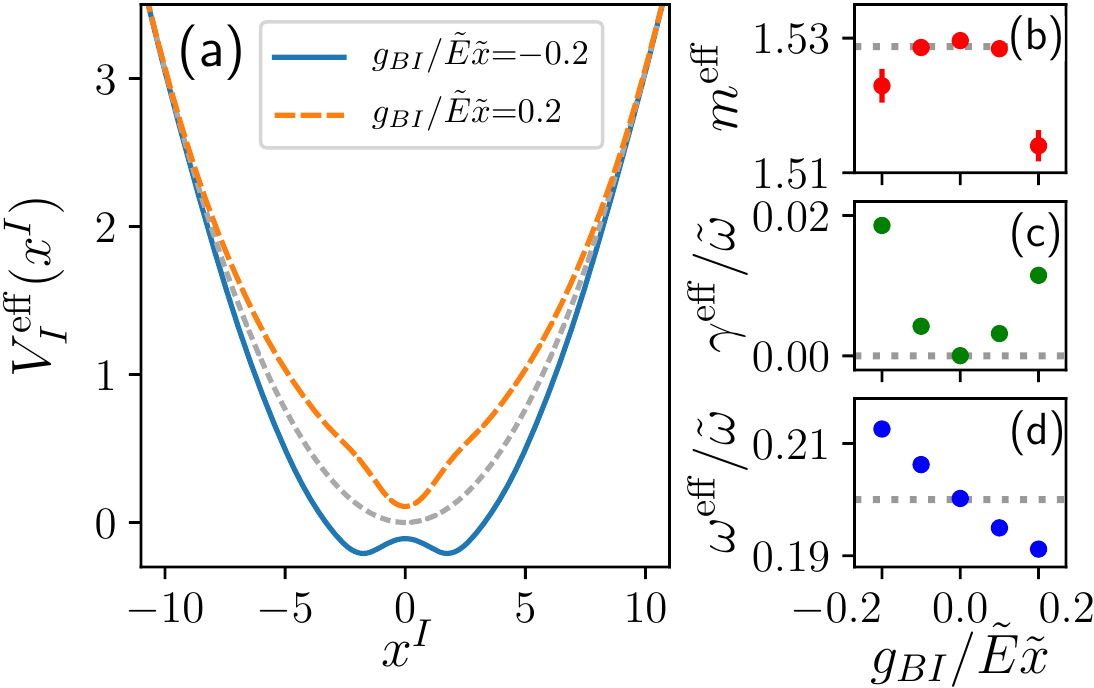}
	\caption{(a) Time-averaged effective potential $V_I^{\textrm{eff}}(x^I)$ [see Eq. (\ref{eq:TAEP})] for $g_{BI}/\tilde{E}\tilde{x}=-0.2$ and $g_{BI}/\tilde{E}\tilde{x}=0.2$ together with the harmonic oscillator potential of the impurity (dashed gray line). The Eqs. (\ref{eq:ap_dho_mp}) and (\ref{eq:ap_dho_mm}) are fitted to the many-body results of the impurity's mean position and mean momentum with respect to the effective mass $m^{\textrm{eff}}$, the damping constant $\gamma^{\textrm{eff}}$ and the effective frequency $\omega^{\textrm{eff}}$, shown in panels (b)-(d), respectively.}
	\label{fig:ap_dho}
\end{figure}

In Figures \ref{fig:ap_dho}(b)-(d) we present the fitted parameters for impurity-medium interactions ranging exemplarily from $g_{BI}/\tilde{E}\tilde{x}=-0.2$ to $0.2$. We find that the effective mass \cite{mistakidis2019c, mistakidis2019a} of the impurity decreases for larger absolute values of $g_{BI}$. This property is attributed to the fact that the bath is confined in a double-well potential. Closely inspecting Figures \ref{fig:gpop}(c1) and (d1) one can observe that a large part of the impurity's density performs a damped oscillatory motion while a smaller fraction of its one-body density accumulates either in the center of the double-well for repulsive impurity-medium interactions or within the double-well sites for attractive couplings.
Consequently, the corresponding dynamics of the mean position and momentum of the impurity capture this damped oscillatory motion. Thereby, the reason of this damping behavior is twofold. Firstly, the non-vanishing impurity-medium coupling and the associated energy transfer \cite{mistakidis2019c, mistakidis2019, nielsen2019} from the impurity to the medium (cf. Figure \ref{ap:energy}) enforce a damped oscillatory behavior on the mean position and momentum. Secondly, the accumulation of the impurity's one-body density around $x^I=0$ additionally enhances the damping of the mean position's and momentum's oscillation, i.e. the decrease of the mean position's and momentum's amplitude in time. Interestingly, the fitting procedure reveals that this damping is not only caused by a damping constant larger than zero, but is also due to an effective mass smaller than the bare value. Furthermore, this effect (damping) can be enhanced by slightly increasing the attractive or repulsive impurity-medium coupling strength (within a parameter range corresponding to the dissipative oscillation regime) leading to a pronounced energy transfer and an accompanied increase of the amount of density accumulated around the trap center. In this sense, we relate the decrease of the effective mass (in the picture of a damped harmonic oscillator) to the accumulation of density around the trap center which reduces the oscillating fraction of the impurity's one-body density. Therefore, the (unexpected) decrease of the effective mass can be traced back to the particular choice of the double-well potential experienced by the bath atoms. Finally, we remark that in the case of a harmonically trapped bath an increase of the effective mass due to the dressing is anticipated \cite{mistakidis2019}.

Moreover, the increase of $\gamma^{\textrm{eff}}$ for increasing attractive and repulsive couplings can be explained by the corresponding growing influence of the bosonic environment. Additionally, we find an approximately linear decrease of the effective frequency $\omega^{\textrm{eff}}$. In order to intuitively explain this behavior we show in Figure \ref{fig:ap_dho}(a) the time-averaged effective potential for the two considered extrema of $g_{BI}$ (i.e. $g_{BI}/\tilde{E}\tilde{x}=-0.2$ and $g_{BI}/\tilde{E}\tilde{x}=0.2$). As can be seen, the effective potential is deeper in the case of attractive $g_{BI}$ compared to the one obtained for repulsive $g_{BI}$ leading to a higher effective frequency $\omega^{\textrm{eff}}$ in the attractive case than in the repulsive one.

Finally, we calculate the effective mass for the impurity in the steady bound state regime for $g_{BI}/\tilde{E}\tilde{x}=-0.9$ [cf. Figure \ref{fig:gpop}(a1)]. Since in this regime the impurity's one-body density exhibits small amplitude oscillations within the left site of the double-well potential the above mentioned procedure can be applied to the mean position and momentum of the impurity. Thereby, we extract an effective mass $m^{\rm{eff}}=1.96\pm0.06$ which is significantly heavier than the bare impurity mass. In contrast to the dissipative oscillation regime in which the accumulation of density around the trap center led to a decreased effective mass, in the steady bound state regime the complete one-body density of the impurity undergoes a damped oscillatory motion which, eventually, leads to the increase of $m^{\rm{eff}}$.

\begin{figure}[t]
	\centering
	\includegraphics[width=1\linewidth]{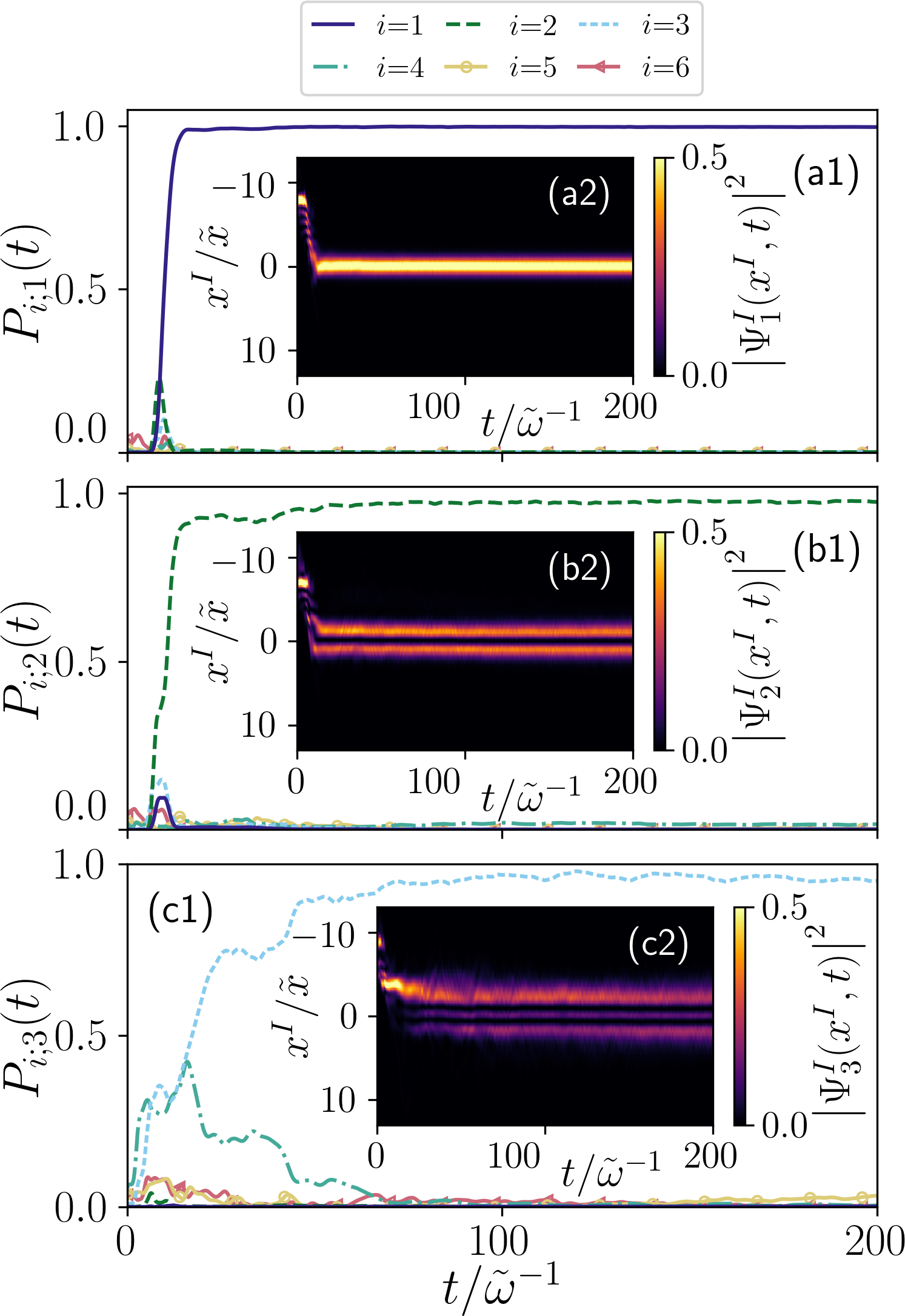}
	\caption{(a1)-(c1) Overlap $P_{i;j}(t) = \rvert\braket{\Psi^I_j(t)} {\psi_i^{\textrm{eff}}}\lvert^2$ of the $j$-th highest occupied species function $\ket{\Psi_j^I}$ of the impurity and the $i$-th eigenfunction of the effective potential [cf. Figure \ref{fig:TAEP_sOccup}(c1)]. Insets (a2)-(c2) show the time-evolution of the respective densities of the species functions. In all panels we set $g_{BI}/\tilde{E}\tilde{x}=1.0$ which corresponds to the pinning regime.}
	\label{fig:ap_speclayer}
\end{figure}

\section{Analyzing the hidden excitations of the pinning regime}
\label{ap:pinning_regime}

The pinning regime appears at intermediate repulsive impurity-medium interaction strengths where the impurity becomes pinned within the bosonic environment residing in the double-well. We attribute the origin of this pinning mechanism to the comparatively large energy transfer of the impurity when it penetrates the bosonic medium for the first time (see Figure \ref{fig:ap_energy}). Furthermore, we have verified that the impurity does not solely occupy the energetically lowest eigenstate of the effective potential, but also populates energetically higher-lying states [Figures \ref{fig:TAEP_sOccup}(c1) and (c2)]. In the main text we referred to these states as hidden excitations since they are not apparent by merely inspecting the evolution of the one-body density $\rho_{I}^{(1)}(x,t)$ [see Figure \ref{fig:gpop}(e1)].

In the following we aim at investigating these hidden excitations in more detail. To this end, we monitor the time-evolution of the density associated with the individual species functions $\ket{\Psi_j^I}$ for $g_{BI}/\tilde{E}\tilde{x}=1.0$ [Figures \ref{fig:ap_speclayer}(a2)-(c2)]. As it can be readily seen, the densities of the species functions remain constant once the impurity is pinned. Moreover, a careful inspection of the densities of the first three species functions reveals an ascending number of nodes which is tantamount to the existence of energetically higher-lying excitations of $\ket{\Psi_j^I}$.

In order to unravel the structure of the aforementioned species functions we project them on the basis set consisting of the eigenfunctions $\psi_i^{\textrm{eff}}$ of the effective potential [cf. Figure \ref{fig:TAEP_sOccup}(c1)] and take the absolute square of the respective overlap, i.e. $P_{i;j}(t) = \rvert\braket{\Psi^I_j(t)}{\psi_i^{\textrm{eff}}}\lvert^2$, where $j=1,2,3$. Indeed, we find that the three dominantly occupied species functions correspond to the three energetically lowest eigenfunctions of the effective potential, see Figures \ref{fig:ap_speclayer}(a1)-(c1). In particular, $|\Psi^I_1(x^I,t)|^2$ matches with the energetically lowest eigenstate $\psi_1^{\textrm{eff}}$, whereas $|\Psi^I_2(x^I,t)|^2$ and $|\Psi^I_3(x^I,t)|^2$ correspond to the second and third eigenstate, i.e. $\psi_2^{\textrm{eff}}$ and $\psi_3^{\textrm{eff}}$.

\section{Dynamical response for a triple-well trapped environment}
\label{ap:triple_well}

To extend our basic conclusions regarding the impurity's response, described in section \ref*{sec:setup}, we replace the double-well of the bosonic medium with a triple-well. However, the harmonic trap of the impurity as well as the employed quench protocol to induce the dynamics remain unchanged.
The triple-well of the bosonic environment reads $V_B(x^B) = m_B\omega_B^2(x^B)^2/2 + g_{-}(x^B) + g_{+}(x^B)$, where a superimposed harmonic trap with frequency $\omega_B/\tilde{\omega}=0.15$ and two Gaussians $g_{\pm}(x^B) = \frac{h_B}{\sqrt{2\pi}w_B} \exp\!\left( -\frac{(x^B\mp\Delta)^2}{2(w_B)^2} \right)$ shifted by $\Delta$ from the trap center are used. Also, the Gaussians have a width of $w_B/\tilde{x}=0.8$ and a height of $h_B/\tilde{E}\tilde{x}=1.8$ while the displacement is $\Delta/\tilde{x}=2.5$.
The system consists of $N_B=20$ bosons for the bosonic medium and a single-impurity $N_I=1$. Additionally, we employ $M=6$ species and $d_{B}=d_{I}=6$ single-particle functions for the calculations to be presented below.

\begin{figure}[t]
	\centering
  	\includegraphics[width=\linewidth]{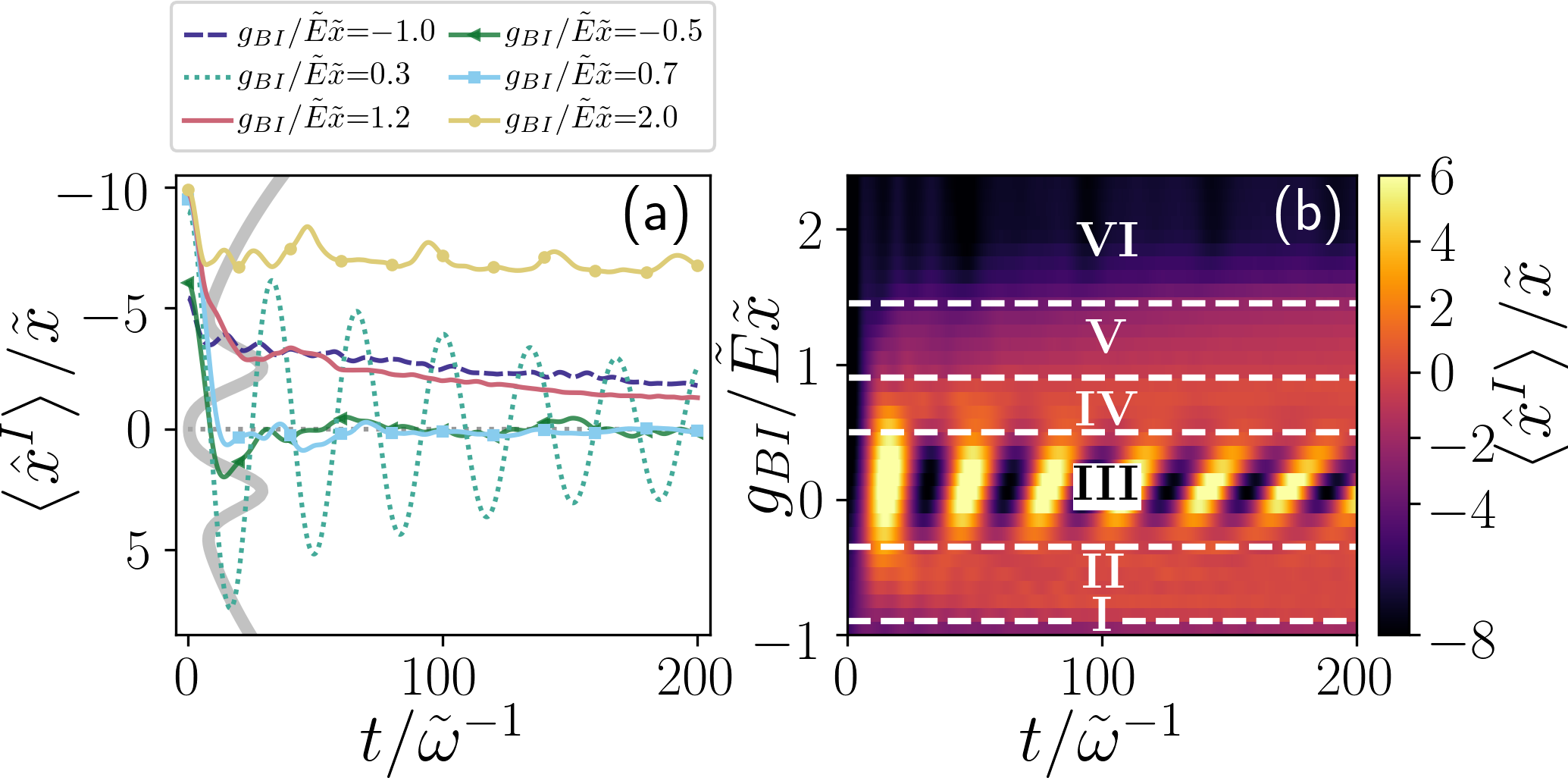}
	\caption{Dynamical response regimes of the impurity as captured by its mean position for the case that the bosonic medium is confined in a triple-well. (a) Time-evolution of the impurity's mean position for specific values of $g_{BI}$ (see legend) corresponding to six identified dynamical response regimes. The triple-well potential of the environment (gray line) is also presented. (b) Dynamical cross-over phase diagram of the impurity's mean position with respect to $g_{BI}$.}
	\label{fig:ap_tw}
\end{figure}

Figure \ref{fig:ap_tw}(a) shows the time-evolution of the impurity's mean position corresponding to the six identified dynamical response regimes which are labeled as I-VI in the respective crossover diagram in terms of $g_{BI}$ illustrated in Figure \ref{fig:ap_tw}(b).
As in the double-well case [see Figure \ref{fig:vNE_mp_cmp}(a)], we find that for strong attractive impurity-medium interactions (regime I) a localization of the impurity in the vicinity of the most left site of the triple-well occurs. This behavior is attributed to the initial large overlap of the impurity with the bath on the left site\footnote{Note that for larger attractions the overlap of the impurity with the medium is initially larger at the central site of the triple-well such that the impurity is initially localized at the trap center where it remains in the course of the evolution.}.
By decreasing $g_{BI}$ to intermediate attractive couplings, i.e. $g_{BI}/\tilde{E}\tilde{x}=-0.5$ corresponding to the regime II, we observe that the impurity localizes at the central site of the triple-well.
Entering the weak attractive and repulsive impurity-medium coupling strengths we identify a dissipative oscillation of the impurity similar to the case in which the bosonic bath is trapped to a double-well [Figure \ref{fig:vNE_mp_cmp}(b)].
For stronger repulsive $g_{BI}$ the aforementioned oscillatory character vanishes (regime IV) and the one-body density of the impurity $\rho_I^{(1)}(x,t)$ (not shown) exhibits two humps located at the two maxima of the triple-well. Here, $\expval{\hat{x}^I(t)}$ tends to zero, e.g. for $g_{BI}/\tilde{E}\tilde{x}=0.7$. Note that even though the regimes II and IV show a similar behavior in terms of $\expval{\hat{x}^I(t)}$ we can distinguish them by evaluating the variance ($s$) which is in the attractive case smaller than in the repulsive one (not shown here).
A further increase of the impurity-bath repulsion to $g_{BI}/\tilde{E}\tilde{x}=1.2$ leads to a localization of the impurity at the position between the left and central site of the triple-well [cf. regime V in Figure \ref{fig:ap_tw}(b)].
For strong repulsive impurity-medium interaction strengths we reach the regime VI which corresponds to the total reflection of the impurity as in the double-well case \cite{theel2020}.

In summary, similarly to the double-well scenario we observe for a triple-well continuous transitions between the emergent dynamical response regimes of the impurity with respect to $g_{BI}$. Analogously to the double-well case, the setup including a triple-well also leads to a bound state and a total reflection regime at strong attractive and repulsive impurity-medium interactions as well as a dissipative oscillation regime at weak $g_{BI}$. Only at intermediate attractive and repulsive $g_{BI}$ corresponding to the effective tunneling and the pinning regime in the double-well case, the impurity features an altered tunneling dynamics. Still, we find that the behavior of the impurity can be steered and controlled via the impurity-medium coupling strength.

\begin{acknowledgements}
	K.K. gratefully acknowledges a scholarship of the Studienstiftung des deutschen Volkes. S.I.M. gratefully acknowledges financial support in the framework of the Lenz-Ising Award of the University of Hamburg. This work has been funded by the Deutsche Forschungsgemeinschaft (DFG, German Research Foundation) --- SFB-925 --- project 170620586.
\end{acknowledgements}

\bibliography{literature.bib}

\end{document}